\documentclass[conference]{IEEEtran}
\IEEEoverridecommandlockouts

\usepackage{tikz}
\usepackage{textcomp}
\usepackage{hyperref}
\usepackage{lipsum}

\newcommand\copyrighttext{%
  \footnotesize \textcopyright 2021 IEEE.  Personal use of this material is permitted.  Permission from IEEE must be obtained for all other uses, in any current or future media, including reprinting/republishing this material for advertising or promotional purposes, creating new collective works, for resale or redistribution to servers or lists, or reuse of any copyrighted component of this work in other works.}
\newcommand\copyrightnotice{%
\begin{tikzpicture}[remember picture,overlay]
\node[anchor=south,yshift=10pt] at (current page.south) {\fbox{\parbox{\dimexpr\textwidth-\fboxsep-\fboxrule\relax}{\copyrighttext}}};
\end{tikzpicture}%
}

\usepackage{cite}
\usepackage{amsmath,amssymb,amsfonts}
\usepackage{graphicx}
\usepackage{booktabs,floatrow}

\usepackage{tabularx}
\usepackage{textcomp}
\usepackage{xcolor}
\usepackage{array,mathtools, multirow}
\usepackage[caption=false]{subfig}
\usepackage{times}
\usepackage{hyperref}
\usepackage{amsmath,amsfonts,bm}

\def\eqref#1{equation~\ref{#1}}
\def\1{\bm{1}}

\def\mD{{\bm{D}}}

\def\mO{{\bm{O}}}

\def\mU{{\bm{U}}}
\def\mV{{\bm{V}}}
\def\mW{{\bm{W}}}
\def\mX{{\bm{X}}}

\DeclareMathAlphabet{\mathsfit}{\encodingdefault}{\sfdefault}{m}{sl}
\SetMathAlphabet{\mathsfit}{bold}{\encodingdefault}{\sfdefault}{bx}{n}

\def\BibTeX{{\rm B\kern-.05em{\sc i\kern-.025em b}\kern-.08em
    T\kern-.1667em\lower.7ex\hbox{E}\kern-.125emX}}

\usepackage{pifont}% http://ctan.org/pkg/pifont
\newcommand{\cmark}{\ding{51}}%
\newcommand{\xmark}{\ding{55}}%

\usepackage{tikz}
\newcommand{\ballnumber}[1]{\tikz[baseline=(myanchor.base)] \node[circle,fill=.,inner sep=1pt] (myanchor) {\color{-.}\bfseries\footnotesize #1};}

\usepackage[capitalise]{cleveref}
    
\begin{document}

\title{NeurObfuscator: A Full-stack Obfuscation Tool to Mitigate Neural Architecture Stealing}
\author{}
\author{\IEEEauthorblockN{Jingtao Li\IEEEauthorrefmark{1}, Zhezhi He\IEEEauthorrefmark{2}, Adnan Siraj Rakin\IEEEauthorrefmark{1}, Deliang Fan\IEEEauthorrefmark{1}, Chaitali Chakrabarti\IEEEauthorrefmark{1}}
\IEEEauthorblockA{\IEEEauthorrefmark{1}School of Electrical Computer and Energy Engineering,
Arizona State University, Tempe, AZ, 85287}
\IEEEauthorblockA{\IEEEauthorrefmark{2}Department of Computer Science and Engineering, 
Shanghai Jiao Tong University, Shanghai}
\IEEEauthorrefmark{1}\{jingtao1, asrakin, dfan, chaitali\}@asu.edu; \IEEEauthorrefmark{2}\{zhezhi.he\}@sjtu.edu.cn
}

\maketitle
\copyrightnotice
\begin{abstract}
Neural network stealing attacks have posed grave threats to neural network model deployment. Such attacks can be launched by extracting neural architecture information, such as layer sequence and dimension parameters, through leaky side-channels.
To mitigate such attacks, we propose  \textit{NeurObfuscator}, a full-stack obfuscation tool to obfuscate the neural network architecture while preserving its functionality with very limited performance overhead. At the heart of this tool is a set of obfuscating knobs, including layer branching, layer widening, selective fusion and schedule pruning, that increase the number of operators, reduce/increase the latency, and number of cache and DRAM accesses.
A genetic algorithm-based approach is adopted to orchestrate the combination of obfuscating knobs to achieve the best obfuscating effect on the layer sequence and dimension parameters so that the architecture information cannot be successfully extracted. 
Results on sequence obfuscation show that the proposed tool obfuscates a ResNet-18 ImageNet model to a totally different architecture~(with 44 layer difference) without affecting its functionality with only 2\% overall latency overhead. For dimension obfuscation, we demonstrate that an example convolution layer with 64 input and 128 output channels can be obfuscated to generate a layer with 207 input and 93 output channels with only a 2\% latency overhead.
% The source code is available at ...
\end{abstract}

\begin{IEEEkeywords}
Neural Network, Side-channel attack, Architecture Stealing, Obfuscation
\end{IEEEkeywords}

\section{Introduction}
The architecture information of a Deep Neural Network~(DNN) model is very sensitive and should never be exposed. It is a valuable Intellectual Property~(IP) that costs companies lots of time and resources.
Knowledge of the exact architecture allows an adversary to build a more precise substitute model and such a model can be used to launch devastating adversarial attacks. For instance, it is shown in~\cite{hu2020deepsniffer} that accurate architecture information enables the adversary to improve the attack success rate of input adversarial attack by almost 3 times. 
% On the other hand, deriving the optimized architecture costs . It is reported that searching an ImageNet archtiecture costs 3150 GPU days using evolution algorithm \cite{real2019regularized}. Thus, it can be considered as a valuable IP.

% Recently, neural architecture search (NAS) technology has been developed to help designers derive neural network architectures that are small, accurate and energy efficient \cite{pham2018efficient, stamoulis2019single, fang2020densely}. 

\begin{figure}[ht]
    \centering
    \includegraphics[width=0.9\linewidth]{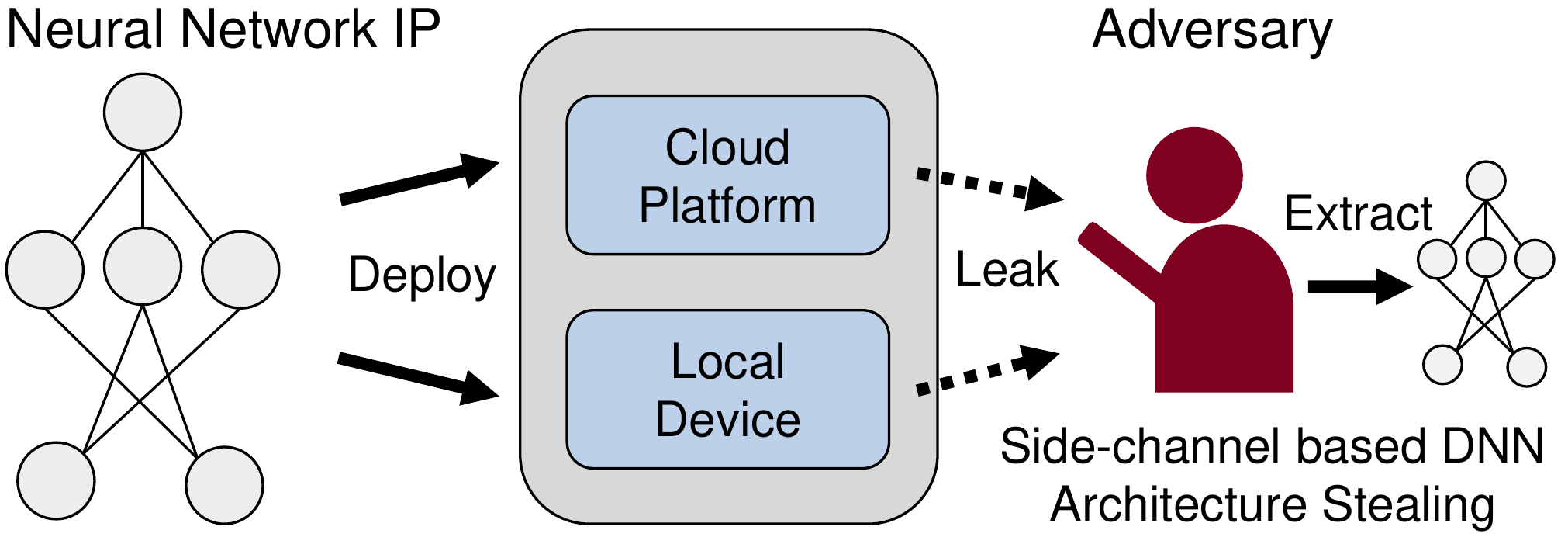}
    \caption{Architecture stealing threatens intellectual property.}
    \label{fig:fig1}
\end{figure}

% However, upon deployment, DNN system can suffer from information leakage. 
%DNN models running on server or local machine cannot avoid being exposed to outsiders either remotely or physically.
Side-channel based DNN architecture stealing has been reported in several prior works~\cite{hu2020deepsniffer, wei2020leaky}. Even without access to the service, an outsider can extract the DNN architecture through side-channel information leakage, as shown in~\cref{fig:fig1}.
Specifically, when the owner of the neural network IP hosts the application on a third-party cloud computing platform or on a local device with GPU support, it opens it up to architecture stealing  through side-channel attacks~\cite{duddu2018stealing, liu2019side, hu2020deepsniffer, wei2020leaky}. A typical architecture stealing flow consists of profiling the target device, training sequence predictor~(e.g., LSTM \cite{sak2014long}), predicting layer sequence based on run-time trace of the target DNN model and then extracting the dimension parameters of each layer.
This is quite different from stealing through Machine-Learning-as-a-Service~(MLaaS) ~\cite{tramer2016stealing, wang2018stealing}, where the attacker has access to the public prediction API and the confidence score of the labels.

%Through remote profiling or local side-channel leakage, the attacker acquire the run-time trace, feed it to extraction neural network and extract the underlying architecture.
% By abusing the hardware profiling program while sitting on the same server, or exploiting side-channel information leakage of a local machine, the attacker can acquire enough knowledge on the underlying computing device, and acquire run-time trace of the target neural network application to extract the entire architecture.

%Thus, architecture IP worth months of development can be stolen/ruined within minutes.

% Even worse, since the black-box condition has been broken, the architecture stealing can leverage more devastating adversary attacks targeting DNN model as has been shown in \cite{carlini2017towards, rakin2019bit}.

%This paper focuses on mitigating DNN architecture stealing that is, extracting the exact architecture topology (layer sequence and dimension parameters) through side-channel information of the target system.

% which focuses more on the approximate functionality while less on the fidelity. It also requires accessing the prediction API to succeed.
%Architecture stealing can be combined with model inversion to increase the fidelity.

Previous efforts on preventing DNN architecture stealing have focused on hardware to eliminate information leakage. Oblivious Random Access Memory (ORAM) technology~\cite{liu2015ghostrider, stefanov2013path} prevents memory access leakage by encrypting the memory address.
\cite{karimi2020hardware} proposes re-design of the Miss Status Holding Registers~(MSHR) to obfuscate GPU memory access to add a layer of randomness.
Though hardware modifications are effective countermeasures, they are not beneficial to existing devices and have high performance overhead. 
Recently, \cite{xu2019gpuguard} proposed a decision tree-based detection method against spy applications on GPU. However, it suffers from high false positive rate and is not practical.
TVM~\cite{chen2018tvm} has also been proposed as a potential countermeasure. Nevertheless, as shown by our experiments~(\cref{fig:tuner_test}), standard TVM does not show enough randomness to be an effective countermeasure.

In this work, we propose \textit{NeurObfuscator}, a full-stack tool which obfuscates neural network execution to effectively mitigate neural architecture stealing. 
Our obfuscating tool consists of 8 obfuscating knobs for two kinds of obfuscation, namely \textit{sequence obfuscation} which obfuscates the layer depth and types and connection topologies between layers,
and \textit{dimension obfuscation} which obfuscates the dimension parameters of each layer, including the number of input and output channel, weight kernel size, etc.
Function-preserving knobs such as layer branching, layer deepening, layer skipping followed by selective fusion at graph optimization step are used for sequence obfuscation, and layer widening, dummy addition, kernel widening and schedule modification in the back end
%AutoTVM operator scheduling\footnote{AutoTVM operator scheduling is a machine learning-based utility in TVM that automatically optimizes the implementation of tensor operators.} 
are used for dimension obfuscation. 

We use genetic algorithm to search for the best set of obfuscation combinations for sequence and dimension obfuscation that achieve strong obfuscation for a given user-defined time budget.  The obfuscation strength is measured by Layer Error Rate~(LER) which represents normalized editing distance of extracted layer sequence given the ground-truth layer sequence in sequence obfuscation, and Dimension Error Rate~(DER) which represents the normalized error of extracted dimension parameters in a layer in dimension obfuscation.
Our contributions can be summarized as follows:
\begin{itemize}
    \item This is the first work on mitigating the NN architecture stealing attack with pure-software obfuscations. We propose a total of 8 obfuscating knobs across the entire DNN execution stack to achieve sequence \& dimension obfuscations and demonstrate the performance on state-of-the-art GPUs.
    
    \item We present an obfuscation tool backed by genetic algorithm to search for the best combination of obfuscations to obfuscate any neural network architecture with user-defined inference latency budget. Source code is available\footnote{Source Code: \href{https://github.com/zlijingtao/Neurobfuscator}{https://github.com/zlijingtao/Neurobfuscator}}.
    
    \item For sequence obfuscation,  our obfuscation tool can obfuscate a ResNet-18 architecture to have a 2.44 LER~(which translates to a 44-layer editing distance \cite{navarro2001guided}) against state-of-art LSTM-based sequence predictors with only 2\% increase in overall latency.
    
    \item For dimension obfuscation, we show how a convolution layer with 64 input and 128 output channels can be obfuscated so that it is extracted as a layer with 207 input and 93 output channels with only 2\% increase in layer-level latency.
     
\end{itemize}

\section{Background}

\subsection{Neural Network Notation}

We summarize the neural network notation that is used throughout the paper, with focus on the most common Conv2D operator~(represented in 4D by $k1,k2,c,j$) in Table~\ref{tab:notation}.

% Weights of a linear operator in layer $i$ are denoted as $\mW_{k, j}^{(i)}$ with shape~($k, j$). The input to the linear operator has shape~($b, j$) and the output has shape~($b, k$), where $b$ denotes the batch size.
% Weights of a Conv2D denoted as $\mW_{k1, k2, c, j}^{(i)}$, is a 4-D tensor of shape~($k1, k2, c, j$), where $k1$, $k2$ are the height and width of the convolution kernel, $c$ is the input channel size and $j$ is the output channel size~(aka. number of filters). The input shape is~($b, c, h_i, w_i$) and the output shape is~($b, j, h_o, w_o$), where $h_i$~($h_o$) and $w_i$~($w_o$) denote the height and width of the input~(output) feature map, respectively.

\begin{table}[ht]
  \caption{Neural Network Notation}
  \label{tab:notation}
  \resizebox{0.8\linewidth}{!}{
  \begin{tabular}{cc}
    \toprule
    Notation  & Definition\\
    \midrule
    $\mX^{(i)}$& Inputs of $i$-th layer \\
     $\mW^{(i)}$, $\mU^{(i)}$, $\mV^{(i)}$ & Weights of $i$-th layer \\
     
    $\varphi(\cdot, \cdot)$& Activation function\\
    $k1$, $k2$& Conv2D kernel sizes\\
    $c, j$ & Input/Output channel sizes\\
    $h_i, w_i$ & Height/Width of inputs\\
    $h_o, w_o$ & Height/Width of outputs\\
    \bottomrule
  \end{tabular}}
\end{table}

\subsection{NN Execution Flow}

\begin{figure}[ht]
    \centering
    \includegraphics[width=\linewidth]{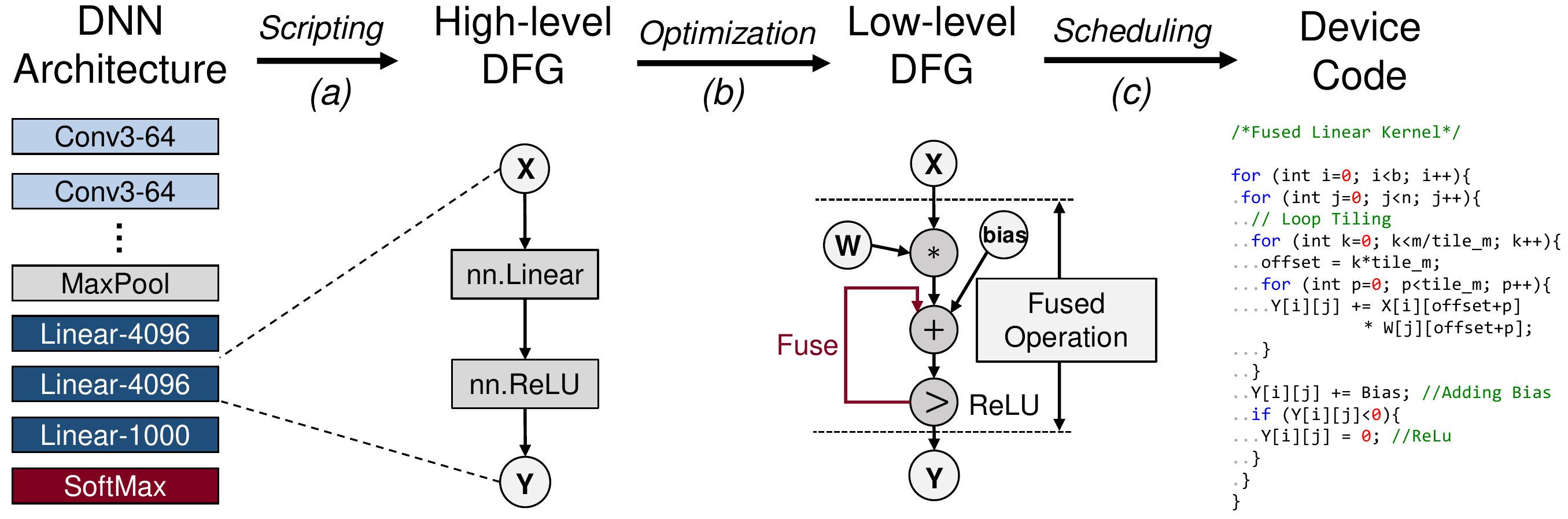}
    \caption{DNN model execution flow. (a) Scripting (Coding) in Python on popular deep learning framework, (b)  using TVM \cite{chen2018tvm}, a deep learning compiler  for graph optimization via TVM Relay, and (c) Auto-TVM scheduling to generate device-level code.}
    \label{fig:fig2}
\end{figure}

% First step is to derive the high-level dataflow graph for the derived NN architecture. This is easy with the help of deep learning frameworks such as Pytorch, Tensorflow. 

Generally, an NN architecture is a topology of neural network layers with non-linear functions. \cref{fig:fig2} demonstrates a typical NN execution process, which consists of multiple steps. 
The first step is scripting~(coding) of a DNN architecture using Python with popular frameworks such as Pytorch or Tensorflow. 
The scripting transforms the raw design into a high-level dataflow graph~(aka. computational graph). 
Next, the high-level graph is ported to TVM for further optimization. 
One can also directly use TorchScript or Tensorflow XLA for graph optimization.
For instance, in TVM, the graph optimization process is handled by Relay module, which provides handy options such as: 1) ``{\tt FoldConstant()}'', which evaluates expression involves only constants; 2) ``{\tt EliminateCommonSubexpr()}'' which creates a shared variable for multiple expressions with same output to avoid the same expression being evaluated multiple times; and 3) ``{\tt FuseOps()}'', which fuses multiple expressions together. User can specify which optimizations to enable.

The last step is scheduling which optimizes the execution of operators on a given device.
% The scheduling is not needed for Pytorch and Tensorflow framework, since they use cuDNN library~(for NVIDIA GPUs) where operators are mapped directly to efficient CUDA codes. 
% The scheduling in Tensorflow is done 
% However, i
In TVM framework, a machine-learning based scheduling called ``AutoTVM''~\cite{chen2018learning} is used to generate optimized codes.
%which has better performance than cuDNN.
For each operator in the optimized low-level graph,  AutoTVM module uses Xgboost~\cite{chen2016xgboost} to search for the best schedule within the predefined search space. \cref{fig:fig2} (c) shows a generic multi-level loop nest implementation of the linear operator. The search space for such a linear operator is defined by one single knob, $tile\_m: [1, m]$, which determines the tiling parameter $m$ for input $X$.

% Even if not using TVM in previous step, users can always import graphs from other frameworks to TVM, which is already a supported feature.

% TVM Relay module is capable of importing high-level dataflow graphs of different framework including ONNX, TensorFlow, PyTorch and Caffe2. Thus, one can write the high-level code in pytorch and port it to TVM to get benefit from graph optimization.

\subsection{Architecture Stealing Attack Flow}
\begin{figure}[ht]
    \centering
    \includegraphics[width=\linewidth]{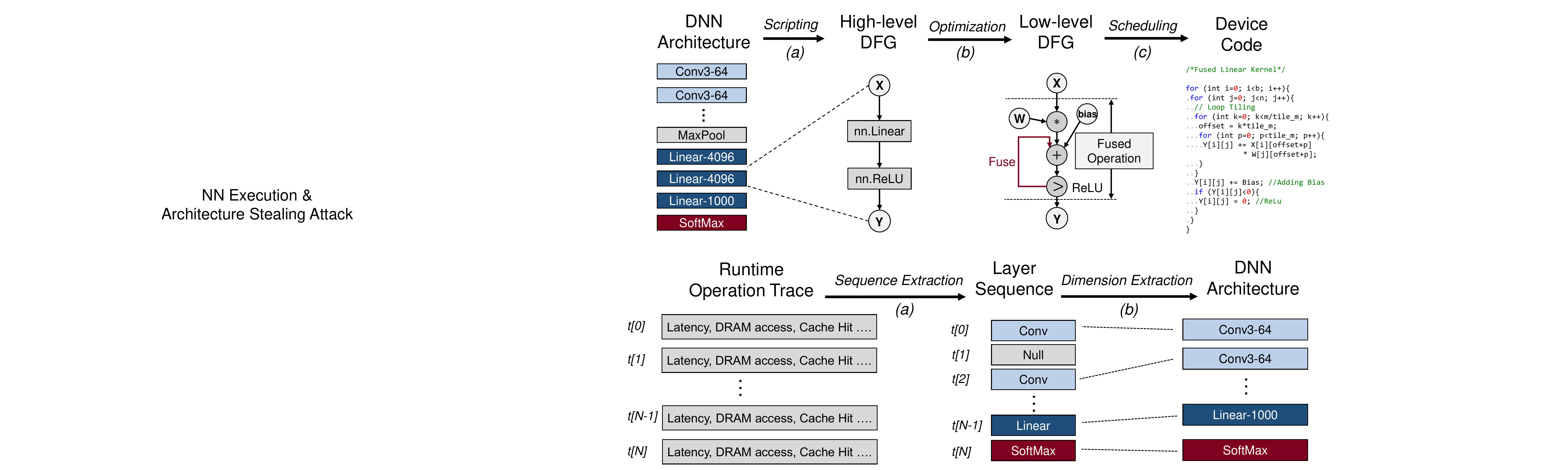}
    \caption{Architecture stealing flow. (a): Layer Sequence Extraction (b): Layer Dimension Extraction.}
    \label{fig:nnflow_rev}
\end{figure}

Extracting the architecture sequence is not trivial. Since neural network execution goes through several steps of optimization as shown in~\cref{fig:fig2}, the intermediate steps bring in lots of variations in the final device code which directly affects the hardware trace.
Prior works~\cite{hu2020deepsniffer, wei2020leaky} have adopted machine learning to extract the architecture from side-channel information. Both works successfully extract common architectures with very high accuracy. They share similar stealing attack methodologies as illustrated in~\cref{fig:nnflow_rev} but differ in their prediction models.

The works in~\cite{hu2020deepsniffer, wei2020leaky} both use Long-Short-Term-Memory~(LSTM) models to predict the layer sequence.
First, massive profiling of randomly generated DNNs on the target devices is done offline. After proper labeling (an example is shown in Fig~\cref{fig:randomNN}), the attacker acquires a trace-sequence dataset and uses it to train the LSTM model. 
At the time of the attack, the attacker uses the LSTM predictor to perform the layer sequence extraction on the run-time trace of the target DNN, which is time-series data consisting of multiple features as shown in~\cref{fig:nnflow_rev}. The sequence prediction locates the \textit{layer operator} in the run-time trace and classifies it by layer type.

Then, dimension extraction is done for each identified layer operator once its  \textit{time-step}~(position) and \textit{class}~(layer type) is known. This is considered to be simpler than sequence extraction. Note that, dimension extraction can be done either manually~\cite{hu2020deepsniffer} or automatically~\cite{wei2020leaky}. 

In summary,  existing architecture stealing attacks heavily rely on the run-time trace, and so
to mitigate such stealing attacks, our obfuscating tool changes the run-time trace as much as possible.

\section{Threat Model}

% We focused on the most common GPU devices as an easy-to-demo example. 
We only consider architecture stealing on applications running on common GPU devices. For other devices such as FPGAs, CPUs and ASICs, we believe that the obfuscation methods proposed in this paper can also be used. We consider NN applications running in both remote and local settings.

% New:

In remote setting,  we assume that the owner runs the NN application on a third-party cloud computing platform and the attacker acts as a normal user without system privilege on the same machine. 
Specifically, the attacker can perform ``driver downgrading attack'' to access the profiling API
%\footnote{Recent NVIDIA driver update has disabled the profiling by default} according to~\cite{wei2020leaky} 
and thus conduct GPU profiling on target neural network applications at run-time. This is similar to the threat model presented in~\cite{wei2020leaky}.

% And at the run-time, attacker access the same cloud service as the target, and use the profiling service to probe target's neural network applications.

In local setting, we assume that the device is off-the-shelf and the attacker can do profiling on an identical device to train a predictor model. While the target application is running, the attacker can get access to the run-time hardware traces of the target neural network applications through side-channel attacks \cite{hu2020deepsniffer, duddu2018stealing, yu2020deepem}.

% Old:

% In remote setting, the attacker can perform ``driver downgrading attack'' to access the profiling API\footnote{Recent NVIDIA driver update has disabled the profiling by default} according to \cite{wei2020leaky}. For models running on local devices, we assume the device is off-the-shelf and the attacker can do profiling on an identical device.

% For neural network applications running on local devices, we assume that the attacker has access to the side-channel information. And for neural network applications running on remote server, we assume that the attacker has access to the the same remote server as the normal user (without system privilege) and has access to the profiling API.

Depending on the attack scenario and capability, we categorize the attacker w.r.t the extent of  information leakage. \cref{tab:cases} describes three cases (from weakest to  strongest):
\begin{itemize}
    \item  \textit{Case-A}: Timing side-channel. The attacker can get accurate operator latency information for each time step. For example, the cycle information for each issued operator is acquired in~\cite{duddu2018stealing, liu2019side}. This case naturally includes Electromagnetic~(EM) side-channel~\cite{yu2020deepem}, as the EM reflects the cycle information of each operator.
    
    \item \textit{Case-B}: DRAM side-channel. The attacker has access to the DRAM read/write information of each operator, as well as the latency through PCIE side-channel~\cite{hu2019neural, hu2020deepsniffer}.
    
    \item \textit{Case-C}: Cache side-channel. The attacker enables context-switching side-channel~\cite{wei2020leaky} or exploits the collocation~\cite{naghibijouybari2018rendered} side-channel. By profiling spy applications, the attacker samples the cache performance counters of the target applications and uses it to extract cache performance, DRAM transactions, latency of the target kernels, etc. The additional cache performance counters in case C include L1 cache and L2 cache utilization, hit rate and read and write data volumes.

\end{itemize}
In all cases, the attacker does massive profiling of DNN model's run-time trace to  steal the architecture.

% , power side-channel \cite{xiang2020open}

\begin{table}[htbp]
\caption{Different Cases of  Information Leakage}
\begin{center}
\centering
% \resizebox{\linewidth}{!}{
\resizebox{0.9\linewidth}{!}{
\begin{tabular}{ccccc} 
 \toprule
%  \textbf{Case} 
 & \textbf{Latency} & \textbf{DRAM-Access} & \textbf{Cache-Counters}  \\ 
 \midrule
 Case-A & \cmark & \xmark & \xmark \\ 
 Case-B & \cmark & \cmark & \xmark \\ 
 Case-C & \cmark & \cmark & \cmark \\ 
 \bottomrule
\end{tabular}}
\label{tab:cases}
\end{center}
\end{table}

\section{Trace Obfuscation}
Architecture stealing is possible because neural network execution process is deterministic, as described in~\cref{fig:fig2}. 
% It is especially the case for popular frameworks (Pytorch and Tensorflow) where graph optimization and compilation steps are deterministic. 
% Only the use of torchscript or tensorflow XLA using JIT-compilation and in TVM framework, the graph optimization process will be added to the noise of the system.
% However, the ``noise'' is not that scary even when scheduling is considered. We found in later Fig.~\ref{fig:tuner_test}, the trace is not very sensitive to the customizable options available in TVM.
To provide countermeasure against architecture stealing, we propose six obfuscating knobs in scripting: layer widening, layer branching, dummy addition, layer deepening, layer skipping and kernel widening. Next, we propose selective fusion under graph optimization and schedule modification in the backend.

% In the rest of this section, we discuss each obfuscating knob in detail.
% The sequence obfuscating knobs (deepen, skip-connection, branching and fission) will be decided first. Because, layer deepen, skip-connection and branching at scripting step will change the searching space of layer fission as they bring more ``complex'' layers. 

\subsection{Obfuscation in Scripting}
%We exploit the large number of  opportunities in the scripting phase of NN execution. 
We realize that many of the function-preserving transformations that have been successfully used in evolution NAS~\cite{zhu2019eena} can be used in obfuscation.
More specifically, we use layer widening, layer branching, layer deepening, layer skipping and kernel widening and dummy addition obfuscating knobs in this phase. 
Note that while many of these operators have been introduced before in the context of architecture evolution~\cite{chen2015net2net, wistuba2018deep}, we are the first to use them in terms of side-channel countermeasures. Layer branching is redesigned, and dummy addition knob is added for dimension obfuscation.

\subsubsection{\textbf{Layer Widening}}
\textit{Layer widening} increases output channel $j$ of a Conv2D layer or a linear layer. Basically, the weights of the added output channels are duplicates of the weights of existing output channels. We allow the widening operator to take fractional numbers. For example, if the weight $\mW_{k_1, k_2, c, j}^{(i)}$ takes a widening factor of 0.25$\times$ and results in $\mU_{k_1, k_2, c, 1.25j}^{(i)}$, then the first $0.5j$ of the output channels come from the duplication of the first $0.25j$ output channel of original $\mW_{k_1, k_2, c, j}^{(i)}$.

To preserve the functionality, next layer's weights need to be adjusted accordingly. In this example, the next layer $\mW_{k_1, k_2, j, m}^{(i+1)}$ must increase its input channel size accordingly, resulting in $\mU_{k_1, k_2, 1.25j, m}^{(i+1)}$ to match the increased output channels. The dimension parameters of $\mU^{(i+1)}$ for the first $0.5j$ input channels have to be adjusted for the duplicated input channels.

\underline{Purpose:} Layer widening increases memory accesses for the current and the next layer by around $(N-1)$ times for widening factor $N$. This results in  increased number of input/output channels and affects dimension extraction.

% \textit{Cost:} Layer widening increases the channel size of the current layer and the next layer, which brings extra $(N-1) \times (c \times j \times k_1 \times k_2 \times h_i \times w_i + j \times f \times k_3 \times k_4 \times h_o \times w_o)$ operators.

\subsubsection{\textbf{Layer Branching}}
\textit{Layer branching} breaks a single NN layer operator into smaller ones. For example, a Conv2D operator $\mW_{k_1, k_2, c, j}^{(i)}$ is branched into two parts (output-wise branching): $\mU_{k_1, k_2, c, j/2}^{(i)}$ and $\mV_{k_1, k_2, c, j/2}^{(i)}$ and the final output is the concatenation of the two partial convolutions:
\begin{equation}
    {\tt Concate}(\mU_{k_1, k_2, c, j/2}^{(i)} * \mX^{(i)}, \mV_{k_1, k_2, c, j/2}^{(i)} * \mX^{(i)})
\end{equation}
While the version in~\cite{zhu2019eena} only considers branching in the output channel dimension of Conv2D/linear layers, we also consider layer branching in the input channel dimension. A Conv2D layer of weight $\mW_{k_1, k_2, c, j}^{(i)}$ is branched into two (input-wise branching): $\mU_{k_1, k_2, c/2, j}^{(i)}$ and $\mV_{k_1, k_2, c/2, j}^{(i)}$, and the final result is the addition of the two:
\begin{equation}
    {\tt Add}(\mU_{k_1, k_2, c/2, j}^{(i)} * \mX^{(i)}, \mV_{k_1, k_2, c/2, j}^{(i)} * \mX^{(i)})
\end{equation}
Here, the activation input needs to be sliced into two as well to match the halved input channel dimension of two smaller convolutions. Various branching methods are feasible, for example, one can also separate it into more than two parts or even do unbalanced branching. Here we consider balanced branching into two or four parts, for both input-wise and output-wise branching.

\underline{Purpose:} 
Layer branching increases the number of layer operators and changes the data volume that needs to be accessed for each operator. For input-wise branching, the input activation and weight volume are halved for each small kernel, and for output-wise branching, input activation is the same but weight and output activation volumes are halved. This knob can be used for both sequence and dimension obfuscation.

\subsubsection{\textbf{Dummy Addition}}
\textit{Dummy addition} is simply adding zero to the activation results. We create a zero matrix of the same shape as the activation output $\mX$ of current layer.
\begin{equation}
    \mD_{b, j, h_o, w_o}^{(i)} = \mO_{b, j, h_o, w_o}
\end{equation}
A dummy addition factor of $N$ means that we create and add the dummy matrix to the output repeatedly  $N\times$.

\underline{Purpose:} Addition operators are ``fused'' into previous layer operators in the fusion step in graph optimization~(refer to~\cref{fig:fig2}) and so the extra cache accesses from the addition operator get added to the layer computation and affect the dimension extraction of that layer.

% \textit{Cost:} Bring extra $N b j h_o w_o$ operators. If the volume is above certain size and cannot fit into the cache, it will result in extra DRAM access, which will increase the latency significantly.

\subsubsection{\textbf{Layer Deepening}}
\textit{Layer deepening} inserts an extra computational layer at the end of current layer's activation function. The insertion of a deepening layer $U^{(i)}$ does not change the original result.
\begin{equation}
    \varphi(\mU^{(i)} * \varphi(\mW^{(i)} * \mX^{(i)})) = \varphi(\mW^{(i)} * \mX^{(i)})
\label{eq:eq4}
\end{equation}
For linear layers, the deepening layer $\mU^{(i)}$ is simply an identity matrix of the same size as its input. For Conv2D layer, layer $\mU^{(i)}$ of size $(k_1, k_2, j, j)$ need to be initialized as:
\begin{equation}
    \mU_{a, b, d, m}^{(i)} =
    \begin{cases}
      0 & \text{$a = \frac{k_1-1}{2} \wedge b = \frac{k_2-1}{2} \wedge d = m$}\\
      1 & \text{otherwise}
    \end{cases} 
\end{equation}
We favor a kernel size of $k_1 = k_2 = 1$ which avoids too much extra computation.
Notice that the correctness of~\cref{eq:eq4} also depends on whether the activation function $\varphi(\cdot)$ results stay the same when it gets stacked $\varphi(\cdot) = \varphi(\varphi(\cdot))$. Fortunately, the most popular ReLU activation subscribes to this property. 
The same property does not hold for batch normalization, so the deepening layer must be added before batch normalization, as shown in~\cref{fig:layer_skipping}~(a).

\underline{Purpose:} Add an extra computational layer to the layer extraction result. This can be used for sequence obfuscation.

% \textit{Cost:} Bring extra $j^2 k_3 k_4 h_o w_o$ operators.

\subsubsection{\textbf{Layer Skipping}}
\begin{figure}[htp]

\includegraphics[width=0.65\linewidth]{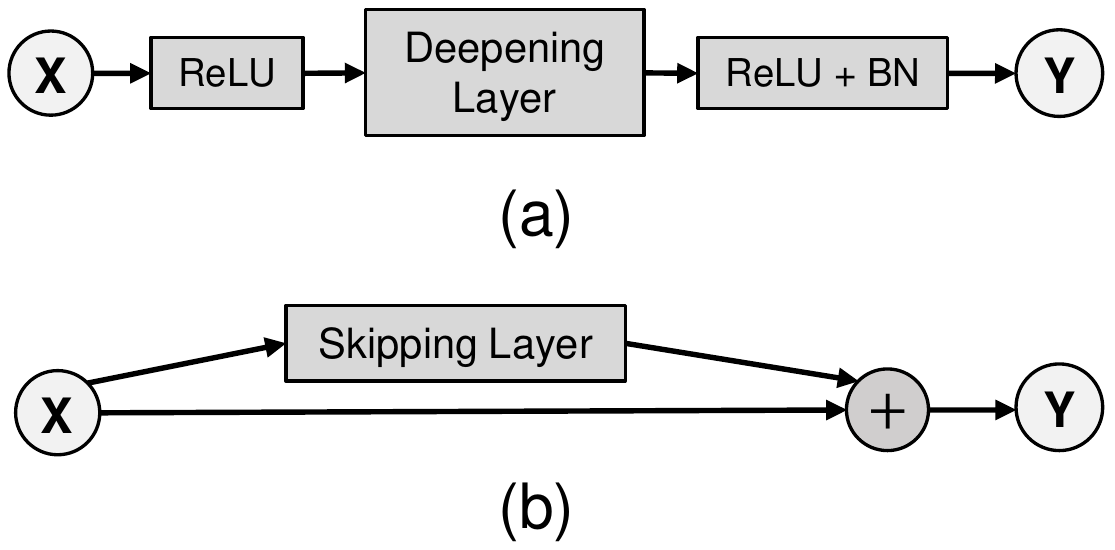}
\caption{(a) Illustration of Layer Deepening. (b) Illustration of Layer Skipping.}
\label{fig:layer_skipping}
\end{figure}

\textit{Layer skipping} inserts an extra computational layer as illustrated in~\cref{fig:layer_skipping}~(b). The additional layer, referred to as {\it skipping layer}, operates on the activation output of an existing layer and adds it to the original activation output. The skipping layer is initialized to zero and thus always have a zero output matrix. 
%To preserve the function, the weights of the skipping layer are initialized to zero.

For an activation of size $(b, j, h_o, w_o)$, the skipping layer can be a Conv2D layer $\mU^{(i)}$ that has a shape of $(k_1, k_2, j, j)$ and all entries are zero. The output of the skipping layer is:
\begin{equation}
    \mU^{(i)} * \mX_{b, j, h_o, w_o} + \mX_{b, j, h_o, w_o} = \mX_{b, j, h_o, w_o}
\end{equation}
    
\underline{Purpose:} Add an extra computational layer to the layer extraction result. This can be used for sequence obfuscation.

% \textit{Cost:} Bring extra $j^2 k_1  k_2 h_o w_o$ computation.    

\subsubsection{\textbf{Kernel Widening}}
\textit{Kernel widening} increases the kernel size of a Conv2D layer. It is done by padding zeros to both the input and convolution kernels. A kernel widening of ``+1'' to a Conv2D layer of shape $(k_1, k_2, c, j)$ will result in new weight of shape $(k_1 + 2, k_1 + 2, c, j)$ and input of shape $(b, c, h_i + 2, w_i + 2)$. We find this to be useful in particular for Conv2D layers that have kernel size of $1 \times 1$. These small $1 \times 1$ kernels would then transform to $3 \times 3$ kernels after widening.

\underline{Purpose:} Change kernel size of the Conv2D operator, resulting in a completely different trace. This  affects dimension extraction.

% \textit{Cost:} A widening of $N$ will bring extra computation:
% \textit{Cost:} A widening of $N$ will bring extra computation:
% $j^2((k_1 + 2N)(k_2 + 2N) - k_1 k_2)h_ow_o$.    

\subsection{Obfuscation in Graph Optimization}
\begin{figure}[htp]

\includegraphics[width=0.96\linewidth]{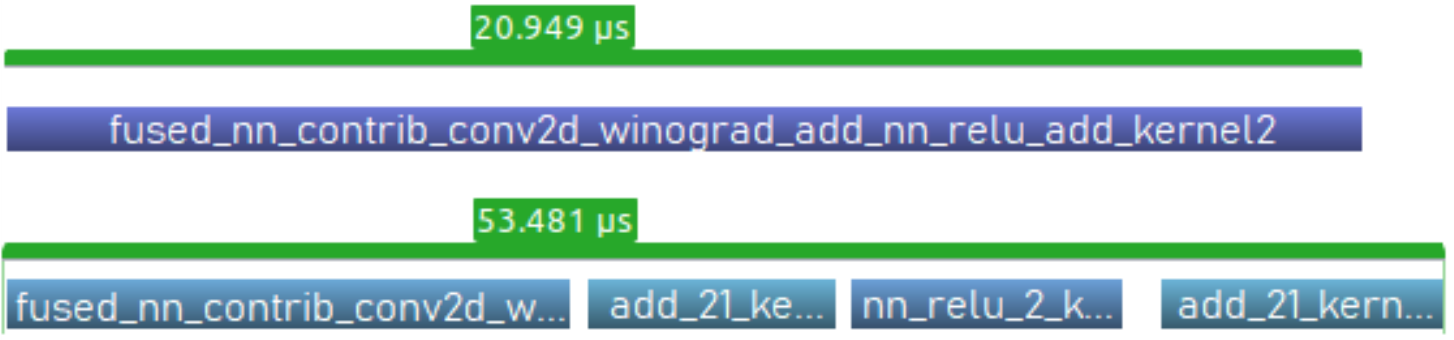}
\caption{Fusion saves time by reducing kernel switching overhead. Top: \textit{winograd\_conv2d\_kernel2} is fused with add and ReLU kernel. Bottom: \textit{winograd\_conv2d\_kernel2} is issued separately from add and ReLU kernels, resulting in significant increase in execution time.}
\label{fig:timeline}
\end{figure}

Fusion is an important graph optimization technique in the TVM Relay module. It fuses subsequent injective operators (scaling or addition) in complex layer operators, such as Conv2D, linear and max-pooling, and transforms the shape of the inputs completely.
% . For example, the subsequent batch normalization and ReLU activation are fused into Conv2D or linear operators and is considered as a single big operator and executed inside a single nested loop.
Fusion ensures execution efficiency as it improves the data reuse and avoids context switching overhead.
% Without fusion, the subsequent batch normalization and ReLU activation are considered as separate operators and will be handled by separate nested loops. 
As shown in~\cref{fig:timeline}, the fused operator is significantly faster than sum of the separate operators.

\textit{7) \textbf{Selective Fusion}:}
\textit{Selective fusion} is a controllable version of the generic  fusion. While the generic fusion fuses successive injective operators greedily, the selective fusion allows $N$ successive operators to fuse and forbids more operators to fuse.
For example, by setting $N$ to zero for a Conv2D operator shown in~\cref{fig:timeline}, the Conv2D operator will be issued separately, as shown in the lower part of the figure.

\underline{Purpose:} Increase the number of operators. Setting $N$ to a small value decreases the memory access and latency of a layer operator, and affects both sequence and dimension extraction.

% \textit{Cost:} Bring extra memory access and kernel switching cost.

\subsection{Obfuscation in Scheduling}

\begin{figure}[ht]
    \centering
    \includegraphics[width=0.9\linewidth]{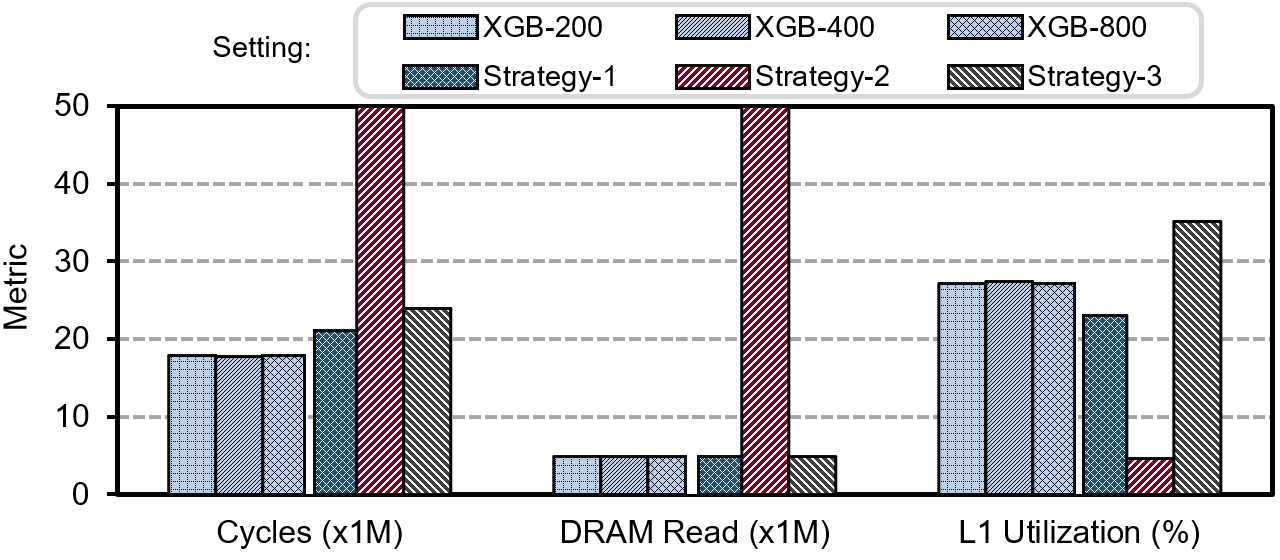}
    \caption{Profiling results of AutoTVM derived schedules using Xgboost tuner with different number of trials (bars 1-3) and  results  using the proposed strategies (bars 4-6).}
    \label{fig:tuner_test}
\end{figure}

In the backend, AutoTVM handles the compilation and generates optimized code for a given device.
It provides options such as the number of trials for tuning, etc. We investigated whether these options can be used to generate randomness in the final result and thereby help in obfuscation. 
We tried 3 rounds with different number of trials using the default Xgboost~(XGB) tuner in AutoTVM for a Conv2D operator. All these trials generated the same schedule, which is understandable because the tuning is designed to optimize latency. The profiling results in~\cref{fig:tuner_test} show that the cycle, DRAM read and L1 cache utilization are very similar for different number of tuning options~(XGB-200, 400 and 800 denoted by bars 1-3), meaning  AutoTVM derived schedule is deterministic and cannot be directly used for obfuscation.

% We tried out different tuner options, and number of trials for the tuning process in AutoTVM module. As shown in Fig.~\ref{fig:tuner_test}, 12 different tuning settings derives only 4 different schedules, distinguished by different ``Tile-Y'' and ``Tile-X'' \footnote{Each entry in Tile-X and Tile-Y decides the tiling size for each dimension of the data, please refer to the topi cuda template for detail.}
% The default xgboost tuner always return the ``option-4'' schedule. In short, only four variation cannot be enough to obfuscate the trace.
% \jtnote{Remaking this figure: focusing on - default usage of xgboost converge to similar solution, and showing some statistics of the one that uses pseudo-prune.}

% Also, changing reward function of the tuning process is not viable, as these counters are not available at the time of tuning.

\textit{8) \textbf{Schedule Modification}:}
To generate schedules via AutoTVM with different outcomes, the search space has to be modified. Actually changing the search space requires a time-consuming tuning (searching) process each time, and so we propose a simple approach that directly modifies the derived schedules with a small sacrifice on operator's performance.
For example, the schedule derived by Xgboost in~\cref{fig:tuner_test} is [-1, 4, 8, 4] for ``Tile-Y'' and [-1, 2, 4, 2] for ``Tile-X''. The first dimension of the tiling is for mini-batch so it is fixed as -1. We present a modification strategy by forcing each of the other dimensions to be 1. For example, the schedule for ``Strategy-1'' forces the second dimension to be 1, which produces [-1, 1, 8, 16] for ``Tile-Y'' and [-1, 1, 4, 4] for ``Tile-X''. We set the values of other two dimensions by keeping the product be the same as original ($8 \times 16 = 4 \times 8 \times 4$) and let them be as close as possible. 
We derived three modified schedules using this method. Their profiling result is shown in bars 4-6 in~\cref{fig:tuner_test}. For latency and L1 cache utilization, all three schedules show noticeable difference. Strategy-2 is an example of bad modification, where the DRAM read and number of cycles explodes and  L1 cache utilization is very poor. Strategy-1, on the other hand, helps achieve obfuscation  without hurting the performance too much.

% A general rule of modifying a schedule is that a strategy that hurts the latency too much should be avoided such as``Pseudo-2''). 

\underline{Purpose:} Derive different schedules for the same operator that present differences in latency, DRAM access and cache performance. This affects dimension extraction.

% \textit{Cost:} Bring extra time cost as the ``pruned'' schedule deviates fro the optimal one.

% \begin{table}[htbp]
% \caption{Summary of obfuscating knobs}
% \begin{center}
% \centering
% \resizebox{\linewidth}{!}{
% \begin{tabular}{ ccccccc } 
%  \toprule
%   & \multicolumn{2}{c}{\textbf{Scripting}} & \multicolumn{2}{c}{\textbf{Graph-Opt.}} & \multicolumn{2}{c}{\textbf{Scheduling}} \\ 
%  \midrule
%  \multirow{2}{*}{\textbf{Sequence}} & deepen & branching$^*$ & \multicolumn{2}{c}{\multirow{2}{*}{fission$^*$}} & \multicolumn{2}{c}{\multirow{2}{*}{$-$}}  \\ 
%  & skip-conn. &  &  &  &  &  \\ 
%  \midrule
%  \textbf{Hyper-} & layer-widen & branching$^*$ & \multicolumn{2}{c}{\multirow{2}{*}{fission$^*$}} & \multicolumn{2}{c}{\multirow{2}{*}{pruning}} \\ 
%  \textbf{Parameter}& kernel-widen & dummy &  &  &  &   \\ 
%  \bottomrule
% \end{tabular}}
% \label{tab:knobs}
% \end{center}
% \end{table}

\section{NeurObfuscator Tool Flow}
\begin{figure}[ht]
    \centering
    \includegraphics[width=0.85\linewidth]{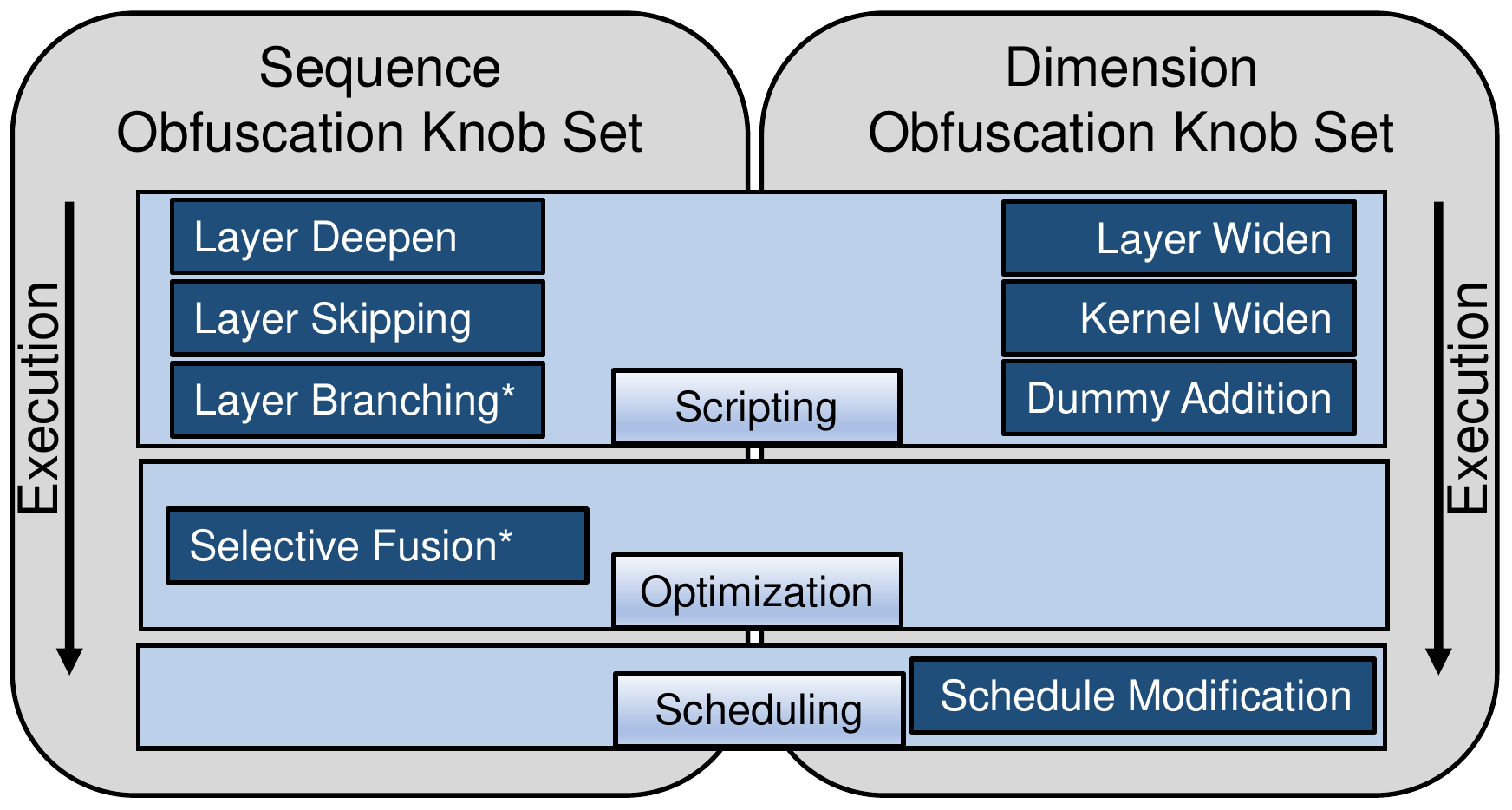}
    \caption{Obfuscating knobs in NeurObfuscator are separated into two sets.}
    \label{fig:knobs}
\end{figure}

The NeurObfuscator tool flow consists of two key steps: 1) sequence obfuscation which obfuscates the layer sequence including layer type and topology, and 2) dimension obfuscation which obfuscates the dimensions of individual layer operators.
We summarize the role of each obfuscating knob in~\cref{fig:knobs}. Knobs with star superscripts
%, such as layer branching and selective fusion, 
affect both sequence and dimension extraction.

% \jtnote{justify the separation of two step [sequence and dimension obfuscation can be taken separately while not affecting getting the global optimum.]}
% In this section, detail for both sequence and dimension is discussed. 
% \subsection{Limiting the Search Space}

\begin{figure*}[htbp!]
    \centering
    \includegraphics[width=0.9\linewidth]{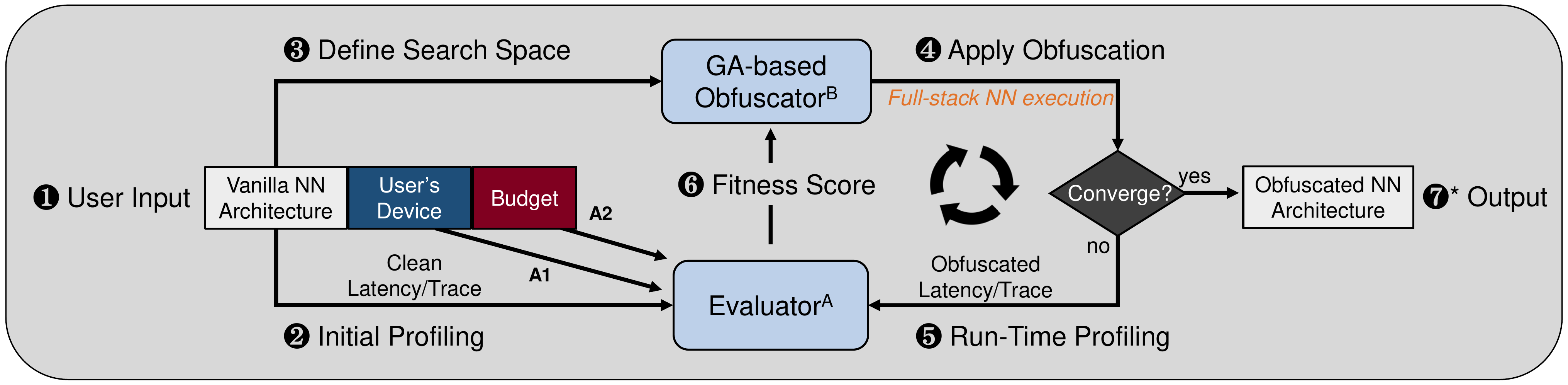}
    \caption{Flowchart overview of proposed NeurObfuscator framework. It takes user's vanilla architecture and outputs an obfuscated NN executable alternative.}
    \label{fig:framework}
\end{figure*}

Since the purpose of sequence and dimension obfuscation is orthogonal, we investigate them independently. 
To reduce search time, we take one more step in limiting the search space as follows.
% However, since the search space for both obfuscations is still huge, we limit the search space in the following way.

\textbf{Knob Partition.} 
First, we partition the set of knobs, as shown in~\cref{fig:knobs}.
We put selective fusion and layer branching together with layer deepening and layer skipping in the sequence obfuscation knob set. The remaining 4 knobs, namely, layer widening, kernel widening, dummy addition and schedule pruning are considered for dimension obfuscation.
Among them, selective fusion and layer branching knobs clearly affect both sequence \& dimension obfuscation. But we exclusively use them in sequence obfuscation where they play a dominant role~(\cref{fig:SQ_individual}).
The four knobs for dimension obfuscation affect dimension parameters significantly while affecting the sequence obfuscation very mildly. In the extreme case, a large change in layer dimension can possibly flip the layer type in sequence obfuscation.

\textbf{Limited Obfuscation Knob Option.} Each obfuscation knob comes with a \textit{list} of options where the $i$-th entry of the \textit{list} denotes a specific obfuscation choice for the $i$-th layer operator. We limit the available options for each entry to reduce the search time. For example, we limit the layer deepening and layer skipping to at most 1, which means at most one deepening layer and one skipping layer can be applied to each layer. 
% For schedule modification, we used ``Strategy-1'' and ``Strategy-3'' static modification strategies, instead of dynamically tuning it on-the-fly.

\textbf{Restricted Search Space.} We restrict the search space by keeping the number of entries~(length of the \textit{list}) for each knob fixed based on the  vanilla architecture.
Otherwise, knobs such as branching, deepening and skipping add extra computational layers and can result in  the search space exploding if they are applied recursively.

\subsection{Sequence Obfuscation}

% \begin{figure*}[htbp!]
%     \centering
%     \includegraphics[width=0.9\linewidth]{Framework.pdf}
%     \caption{Frame overview of proposed NeurObfuscator. It takes user's vanilla architecture and outputs an obfuscated NN executable alternative.}
%     \label{fig:framework}
% \end{figure*}

To derive the best set of obfuscating knobs for sequence obfuscation,
% . Because of their sequence obfuscating effect and latency increase are hard to quantify because of the complicated execution stack, 
we model this as a combinatorial optimization problem and solve it using genetic algorithm. Basically we find the set of sequence obfuscating knobs such that the obfuscated NN achieves strong obfuscation and can be executed within a given  time budget.
The obfuscation metric is given by layer prediction error rate or LER and the time budget is a small fraction of the inference latency. The overview of the obfuscation framework is given in~\cref{fig:framework}.

The input to the framework is the vanilla model and time budget (steps \ballnumber{1}-\ballnumber{3}). The device where the framework is running on is also an underlying input since it determines the trace. As discussed previously, the search space is  based on the vanilla architecture.
An initial profiling is done to derive clean latency $T^*$ and clean trace.
%which are used for calculating the fitness score.
The obfuscator applies a selected set of obfuscation knobs in step~\ballnumber{4} and runs inference. The profiling is done on the obfuscated model in step~\ballnumber{5} and the evaluator calculates a fitness score given the latency and layer prediction error rate (LER)  in step~\ballnumber{6}.
When fitness score converges (circling through steps \ballnumber{4}, \ballnumber{5}, \ballnumber{6}), the framework outputs the compiled binaries of the obfuscated model.

% The optimization process is shown on the left, where the obfuscator picks a obfuscation and applies on the vanilla NN architecture. Then, the evaluator returns fitness score given trace and the inference latency of the obfuscated neural network. This process is iterated over and finally gives the obfuscation with the highest score.

\subsubsection{Evaluator: LSTM Predictor Testbed}

To evaluate the obfuscation effect, we build a testbed that performs stealing attack on the obfuscated architecture based on existing stealing methods in~\cite{hu2020deepsniffer, wei2020leaky}.

\begin{figure}[htbp]
    \centering
    \includegraphics[width=\linewidth]{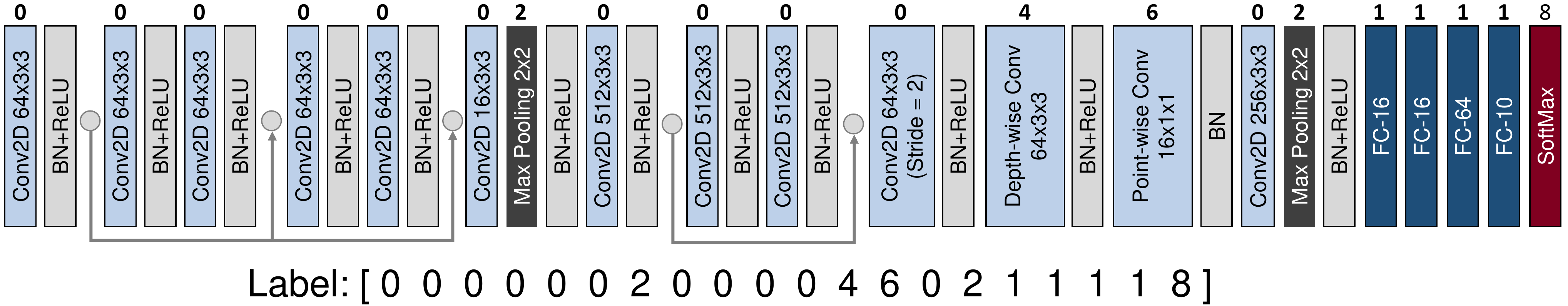}
    \caption{Random NN archtecture and its labeling. Labeling is only done on Conv2D, Linear, MaxPool and SoftMax layers which are considered as complex layer operators and will not be fused.}
    \label{fig:randomNN}
\end{figure}

\textbf{Dataset Generation.} To mimic the attacker, first, massive profiling on the user's device needs to be done. So we build a random neural network architecture generator, which is used as input to the profiling toolset. 
It first fixes the depth of the network~(number of computational layers), and at each step,  randomly inserts neural network convolution layer with random dimension parameters~(input channel size and output channel size), ResNet and MobileNet computing blocks and pooling/batch normalization (BN) layers. 
 Linear layers with random number of neurons are added only after all the Conv2D layers. The classification layer~(linear layer with neuron equals to the number of class) and the softmax layer are added at the end. 
We generate 6,000 different neural network architectures for input size of [3, 32, 32]  and number of classes equals to 10, to match the CIFAR-10 dataset setting. We generate another 6,000 architectures for input size of [3, 224, 224], and number of classes equal to 1,000 to match the ImageNet dataset setting.
Because normally the BN/ReLU are fused with complex layer operators (Conv2D, Linear, etc.), we only label the complex operators. An example of the randomly generated architecture is shown in~\cref{fig:randomNN}.

\textbf{Run-Time Profiling.} Both the offline and run-time profiling is done using Nsight Compute\footnote{A proprietary tool maintained by NVIDIA corp. for CUDA kernel profiling, similar to NVPROF.}, which uses ``kernel replay'' for accurate trace generation.
We use this tool to simulate the three cases (cases A, B and C) of attack described in Section III.
Two contemporary NVIDIA GPUs are used for profiling, i.e., a Turing GPU~(GTX-1660) to profile models on CIFAR-10 dataset and an Ampere GPU~(RTX-3090) to profile models on ImageNet dataset. We collect the number of cycles, DRAM and cache performance metrics for each issued operator of the model running in inference mode. After profiling, we collect three sets of features to match the three attack cases. The selected features in Nsight Compute profiling process are listed in~\cref{tab:feature_collection}. 
In practice, the attacker gets noisy trace information through side-channels~\cite{wei2020leaky, hu2020deepsniffer}.
To study the worst-case~(i.e., strongest attack), we assume that the attacker that can obtain an accurate trace.

\begin{table}[htbp]
\caption{Feature measurement enabled in profiling}
\begin{center}
\centering
\resizebox{\linewidth}{!}{
\begin{tabular}{ lr} 
 \toprule
  \textbf{Feature} & exists in \textbf{Case}\\ 
 \midrule
 sm\_\_cycles\_active.sum & A, B, C\\ 
 \hline
 dram\_\_sectors\_read.sum, dram\_\_sectors\_write.sum & B, C\\ 
 \hline
 L1 transaction\footnote{l1tex\_\_t\_sectors\_pipe\_lsu\_mem\_global\_op\_ld/st.sum}, utilization\footnote{l1tex\_\_lsu\_writeback\_active.avg.pct\_of\_peak\_sustained\_active}, hit rate\footnote{l1tex\_\_t\_sector\_hit\_rate.pct}& \multirow{2}{*}{C}\\
 L2 transaction\footnote{lts\_\_t\_sectors\_op\_read/write.sum}, utilization\footnote{lts\_\_t\_sectors.avg.pct\_of\_peak\_sustained\_elapsed}, hit rate\footnote{lts\_\_t\_sector\_hit\_rate.pct}& \\
 \bottomrule
\end{tabular}}
\label{tab:feature_collection}
\end{center}
\end{table}

% Traces have many-to-one mapping to actual operators (i.e. several traces can represent one Conv2D layer). Thus, CTC loss is a effective technique as it provides Null labels to classify time-steps that are trivial, which is adopted in \cite{hu2020deepsniffer}.

% and the prediction problem setting is very similar to OCR and automatic speech recognition. Thus, CTC loss is a effective technique as it provides Null labels to classify time-steps that are trivial, which is adopted in \cite{hu2020deepsniffer}.

\textbf{LER metric.} The LSTM-based predictor for the testbed is a single-layer LSTM-RNN model with a Connectionist Temporal Classification~(CTC) decoder as adopted in Deepsniffer~\cite{hu2020deepsniffer}. The \textit{Layer prediction Error Rate}~(LER) is used to quantitatively measure the performance of a trained predictor.
The LER has the form:
\begin{equation}
    LER = \frac{ED(L, L^*)}{|L^*|}
\end{equation}
where $L$ is the predicted sequence and $L^*$ is the ground-truth, $ED$ denotes editing distance~(Levenshtein distance~\cite{navarro2001guided}) and $|\cdot|$ denotes the length.

We derive three sets of LSTM predictors - one for each attack case. For each case, we set different number of hidden units of the LSTM network to 64, 96, 128, 256 and 512, resulting in a total of 3$\times$5 = 15 LSTM predictors.
We split each dataset into 4:1 for training and validation subsets, and train for 150 epochs. The final validation LERs for all the LSTM predictors are shown in~\cref{tab:predictors}. We observe an excellent layer sequence extraction performance on case C where all the latency, DRAM and cache features are considered for each time-step, and a comparative poor performance on case A where only the latency feature is considered.

\begin{table}[htbp]
\caption{Validation LER of LSTM-based layer Sequence Predictors}
\begin{center}
\centering
\resizebox{\linewidth}{!}{
\begin{tabular}{ cccccccc } 
 \toprule
 \multirow{2}{*}{\textbf{LSTM unit}} &
%  \textbf{LER} &
 \multicolumn{3}{c}{\textbf{CIFAR-10}} & \multicolumn{3}{c}{\textbf{ImageNet}}  \\ 
 \cmidrule(r){2-4}\cmidrule(r){5-7}
%  \cline{2-7}
  & case A & case B & case C & case A & case B & case C  \\ 
 \midrule
 64-unit & 0.095 & 0.100 & 0.001 & 0.178 & 0.027 & 0.002  \\ 
 96-unit & 0.121 & 0.087 & 0.007 & 0.291 & 0.035 & 0.000  \\ 
 128-unit & 0.126 & 0.045 & 0.008 & 0.292 & 0.044 & 0.001  \\ 
 256-unit & 0.077 & 0.023 & 0.013 & 0.283 & 0.031 & 0.004  \\ 
 512-unit & 0.098 & 0.074 & 0.000 & 0.303 & 0.036 & 0.003  \\ 
 \bottomrule
\end{tabular}}
\label{tab:predictors}
\end{center}
\end{table}

\begin{figure}[htbp]
    \centering
    \includegraphics[width=0.85\linewidth]{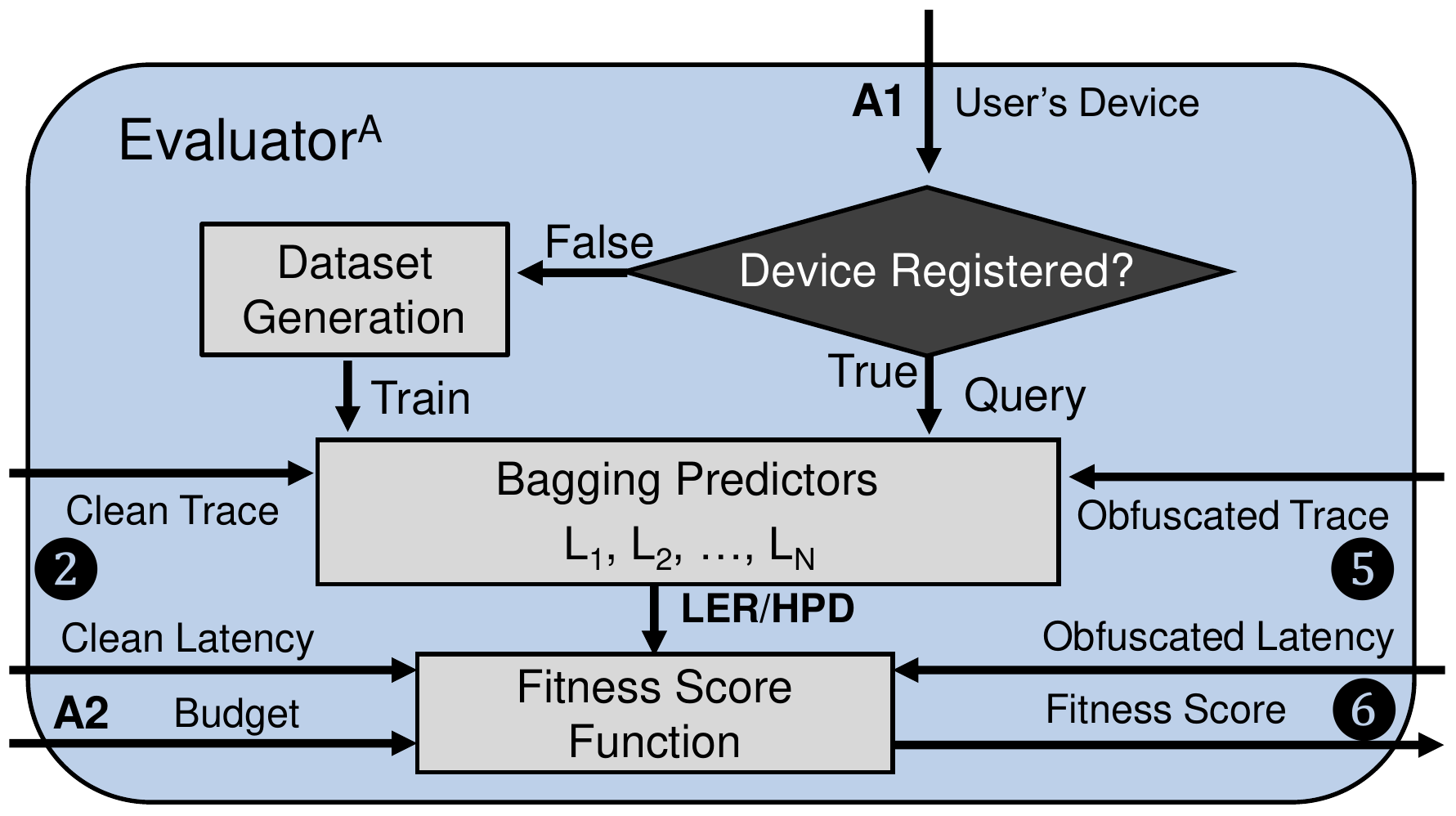}
    \caption{Design of the \textbf{Evaluator} in the proposed framework. Bagging predictors supply LER (in sequence obfuscation) or DER (in dimension obfuscation) to calculate the fitness score function.}
    \label{fig:evaluator}
\end{figure}

\textbf{Predictor Training.} The evaluator uses the bagging approach \cite{breiman1996bagging} and provides the average LER of LSTM predictors for different input sizes, where we choose the input sizes to match that of CIFAR-10 and ImageNet datasets. 
% To evaluate the obfuscation, we use bagging of all predictors to give an average LER score on the obfuscated model, as we cannot anticipate the capability of the attacker (case A, B or C). 
The evaluator is shown in~\cref{fig:evaluator}. The training needs to be done once for each new device.
% We provide trained LSTM for two types of GPUs (RTX-3090 and GTX-1660) 
% The predictor training can be reused for GPUs having the same type (i.e. GTX-1660), which can potentially save lots of time.  

\subsubsection{GA-based Obfuscator}
% \textbf{Fitness Score:}
Our next goal is to maximize the obfuscation given a user-defined latency budget, $B$.
 For instance,  $B = 0.1$ means that the  user can afford up to 10\% extra inference latency.  Then the optimization problem can be set up as a constrained discrete optimization problem that maximizes the average LER given the latency budget: 
\begin{equation}
\begin{aligned}
\min_{S} \quad & \frac{1}{N}\sum_{i=1}^{N}{LER_{i}(S)}\\
\textrm{s.t.} \quad & T\leq (1 + B) T^*\\
\end{aligned}
\label{eq:fitness1}
\end{equation}
where, $N$ is the number of predictors in bagging, $S$ denotes the set of obfuscation options,
% with discrete values {\bf corresponding to settings of?} each obfuscating knob list. 
$T$ is the latency with obfuscation and $T^*$ is the clean latency without obfuscation.

% The obfuscation set has 8 components corresponding to 8 obfuscating operators described in section V. We use python dictionary object to construct the obfuscation set. We represent each item of the set as a list. For example, the key `widen\_list' is a list has the size as the total number of neural network layer, and each entry denotes the widening factor ($\geq$ 1.0). The `deepen\_list' and `skipcon\_list' are of the same length and each entry is binary denoting whether to add an extra layer. A detailed description of the API is available in the documentary of the open-sourced repository.
% \jtnote{For the full description, Please see the anonymous repository}

\begin{figure}[htbp]
    \centering
    \includegraphics[width=0.85\linewidth]{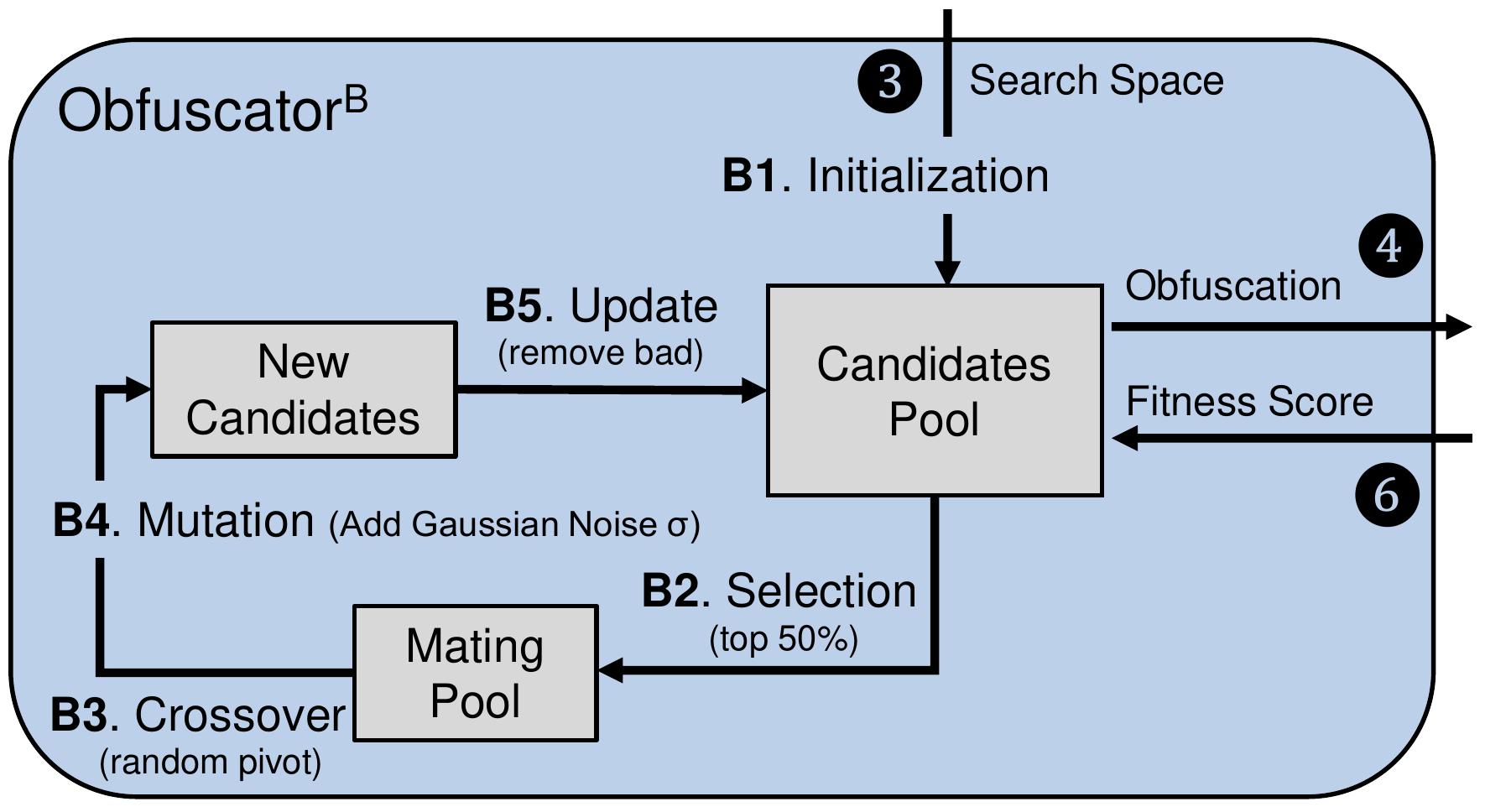}
    \caption{Design of the GA-based \textbf{Obfuscator} in the proposed framework.}
    \label{fig:obfuscator}
\end{figure}

\textbf{Genetic Algorithm.}
We choose to use the genetic algorithm (GA) to solve the discrete optimization problem. Since the optimization with \textit{constraints} in~\cref{eq:fitness1} cannot be directly used in GA (as GA works well for unconstrained optimization problems~\cite{rajeev1992discrete}), the reward $R$ (a.k.a. fitness score) for GA is designed as follows:
\begin{equation}
% \begin{array}{rclcl}
\displaystyle R = \frac{1}{N}\sum_{i=1}^{N}LER_{i}(S) \,\bigg/\, \biggl[\epsilon + \biggl(\dfrac{T - (1 + B) T^* }{T^*}\biggr) ^2 \biggr]
% \end{array}
\label{eq:fitness2}
\end{equation}
We replace the constraints in~\cref{eq:fitness1} with a penalty term, which penalizing the reward  when latency $T$ deviates from the total latency $(1 + B) T^*$. This deviation is normalized and squared and a small offset term $\epsilon$ is added to avoid zero proximity.
The block diagram of GA-based obfuscator is shown in~\cref{fig:obfuscator}.

The initial value of each obfuscating knob is randomly generated based on the search space provided in step~\ballnumber{3} in~\cref{fig:framework}.
For the mating process, we add top 50\% of the population based on fitness score into the mating pool. The crossover process takes random pivot of two lists and produces the same number of offsprings. In the mutation process, we apply Gaussian noise with standard deviation $\sigma$ on each of the offspring obfuscation sets with rounding and clipping to keep the value in legal format. Mutated offsprings are added into the candidate pool, and half of the candidates (including newly added offsprings) that have lowest fitness scores are removed from the candidate pool. The pool of candidates gradually improve over generations resulting in high fitness scores. 

\subsection{Dimension Parameter Extraction}

For dimension obfuscation, we focus on the obfuscating knobs, namely, layer widening, kernel widening, dummy addition and schedule modification, that affect the dimension parameters the most. We only describe obfuscation on standard Conv2D operators as they appear most frequently in the DNN architectures under testing.

\textbf{DER metric.} To evaluate the prediction error of a layer's dimension parameter, we define dimension parameter prediction error rate (DER) as a measure of the obfuscation effect similar to the LER metric. If the number of input/output channels of a Conv2D operator
be $(c, j)$, the DER for a given prediction $(c, j)$ on layer $i$ is defined as:
\begin{equation}
    DER(i) = \frac{|c - c^*|}{c^*} + \frac{|j - j^*|}{j^*}
    \label{eq:DER}
\end{equation}
where, $c^*$ and $j^*$ represent the original (without obfuscation input/output channels.

\textbf{Predictor Training.} We adopt the Random Forest~(RF) model, as a bagging version of decision trees for the dimension parameter  extraction testbed. We collected around 50,000 traces for Conv2D operators with different input channel and output channel parameters ($c$, $j$), which are the two most important dimension parameters. We neglect stride, kernel size and padding features because they rarely change.
We train RF regression model with different number of trees (30, 50, 100 and 200) to predict $c$ and $j$ separately. The training and validation ratio is set to 4:1. We record the average DER of the validation dataset (20\% of the data) for ImageNet and CIFAR-10 for the three attack cases. The results in~\cref{tab:randomforests} show that the dimension extraction has negligible error for cases B and C and comparably high error for case A because it has only latency feature.
Furthermore, the number of trees do not affect the prediction performance much.

% \begin{table}[htbp]
% \caption{MAE of Random Forest regression model for Dimension Extraction.}
% \begin{center}
% \centering
% \resizebox{0.95\linewidth}{!}{
% \begin{tabular}{ cccccccc } 
%  \toprule
%  \multirow{2}{*}{\textbf{MAE}} & \multicolumn{3}{c}{\textbf{CIFAR-10}} & \multicolumn{3}{c}{\textbf{ImageNet}}  \\ 
%  \cmidrule(r){2-4}\cmidrule(r){5-7}
%   & A & B & C & A & B & C  \\ 
%  \midrule
%  Input Channel $c$ & 21.79 & 1.88 & 0.022 & 6.90 & 1.14 & 0.91  \\ 
%  \midrule
%  Output Channel $j$ & 23.95 & 5.45 & 0.005 & 13.92 & 1.46 & 1.28  \\ 
% %  \midrule
% %  Stride & 0.027 & 0.000 & 0.000 & 0.007 & 0.0007 & 0.0004  \\
%  \bottomrule
% \end{tabular}
% }
% \label{tab:randomforests}
% \end{center}
% \end{table}

\begin{table}[htbp]
\caption{Average DER of Random Forest regression model for Dimension Extraction.}
\begin{center}
\centering
\resizebox{\linewidth}{!}{
\begin{tabular}{ cccccccc } 
 \toprule
 \multirow{2}{*}{\textbf{Number of Trees}} & \multicolumn{3}{c}{\textbf{CIFAR-10}} & \multicolumn{3}{c}{\textbf{ImageNet}}  \\ 
 \cmidrule(r){2-4}\cmidrule(r){5-7}
  & case A & case B & case C & case A & case B & case C  \\ 
 \midrule
 30-tree &  0.467 & 0.060 & 0.014& 0.160 & 0.041 & 0.023  \\ 
 50-tree &  0.465 & 0.060 & 0.014& 0.160 & 0.041 & 0.023  \\ 
 100-tree & 0.462 & 0.059 & 0.014& 0.160 & 0.040 & 0.023  \\ 
 200-tree & 0.464 & 0.059 & 0.014& 0.159 & 0.040 & 0.023  \\ 
%  \midrule
%  Stride & 0.027 & 0.000 & 0.000 & 0.007 & 0.0007 & 0.0004  \\
 \bottomrule
\end{tabular}
}
\label{tab:randomforests}
\end{center}
\vspace{-5pt}
\end{table}

The dimension obfuscation framework is similar to that of sequence obfuscation framework shown in~\cref{fig:framework}. 
% It starts with the NN architecture that has gone through sequence obfuscation. 
The user needs to specify a budget that limits the latency increase in dimension obfuscation for each layer.
The evaluator (\cref{fig:evaluator}) uses RF regression model as the ``Bagging Predictor'' and average DER of all three attack cases to compute the fitness score.
Note that the GA-based obfuscator (\cref{fig:obfuscator}) for dimension obfuscation has a different search space because a different set of obfuscating knobs is considered.

\section{Evaluation}

\subsection{Sequence Obfuscation Performance}
% \jtnote{Add more analysis to each subsection}
We evaluate the performance of our obfuscation tool on a series of standard models \cite{simonyan2014very, he2016deep, howard2017mobilenets}. Specifically, we select VGG-11, VGG-13, ResNet-20, ResNet-32 models on CIFAR-10 running on a Turing GPU (GTX-1660). We select VGG-19, ResNet-18 and MobileNet-V2 models on ImageNet running on an Ampere GPU (RTX-3090). 
% In addition, we also include a model from the validation dataset that consists of 19 computational layers includes two ResNet BasicBlocks and one MobileNet Computing Block, denoted as ``Custom-NN''. 
For the GA, we set the population size to be 16 and run it till the fitness score stabilizes which occurs around 20 generations. The standard deviation $\sigma$ for the the mutation step is set to a high value (i.e. $\sigma = 8.0$) at the beginning and gets halved after every 4 generations. 
%Under this GA setting, we observe that the solution consistently converges for different runs. 
To eliminate the randomness, for each data point reported here, we choose the average of 3 runs.

\textbf{Effect of Individual Knobs}. 
% On a convolution layer of 64 input channel and 128 output channel (second layer of VGG-11), 
First, we investigate the effect of individual knobs on  stand-alone Conv2D operators with different dimension parameters. We list the latency overhead for 
different dimension parameters in Table~\ref{tab:individual_knobs}.  
We found layer branching introduces extra operators with a low latency cost. For example,  output-wise layer branching into four adds 3 extra Conv2D operators and 1 concatenate operator with at most 49\% latency increase. Selective fusion increases latency  by around 15\% but it only introduces one extra ReLU operator and BN operator. In contrast, deepening layer and skipping layer introduce an extra Conv2D operator at a lot higher latency cost, and is thus not effective.

\begin{figure}[htbp]
    \centering
    \includegraphics[width=1.0\linewidth]{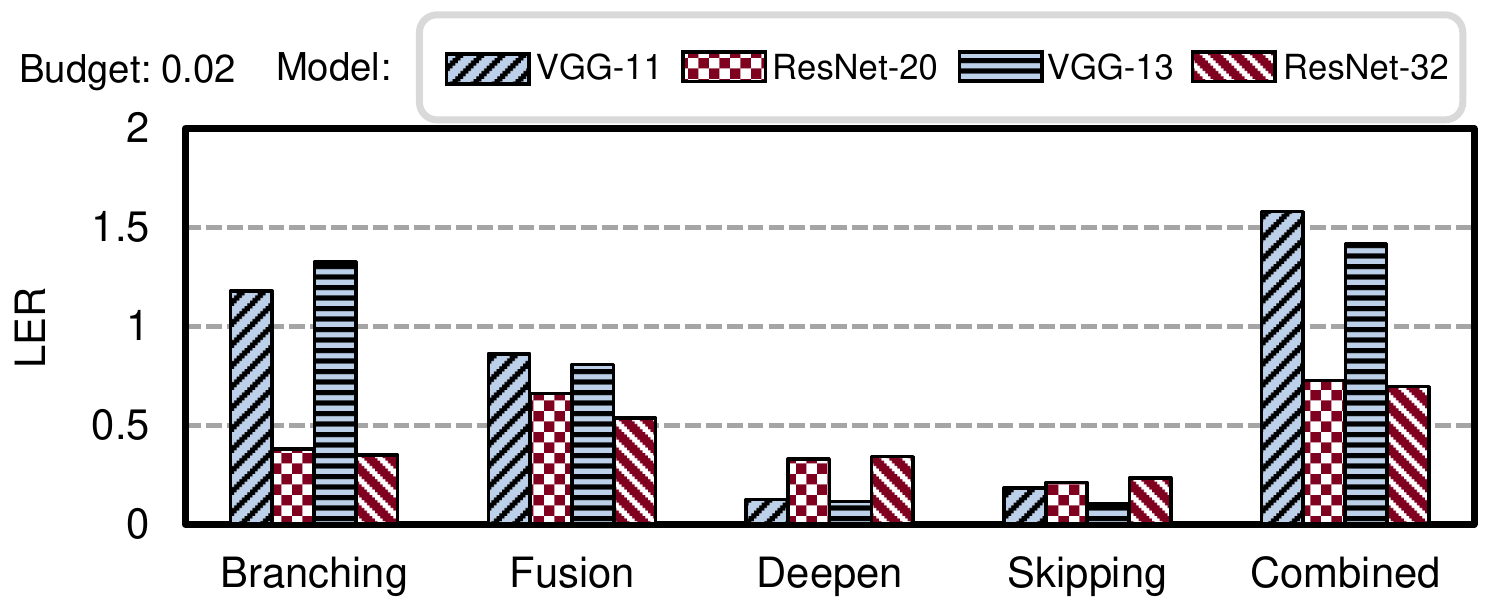}
    \caption{Individual contribution of each sequence obfuscating knob  in GA-obfuscator, followed by combination effect. Time budget is selected at 0.02. VGG-11, VGG-13, ResNet-20 and ResNet-32 are on CIFAR-10 dataset.}
    \label{fig:SQ_individual}
\end{figure}

Since the latency overhead due to application of an obfuscation knob on a single operator is large, the obfuscation knobs have to be applied selectively to only certain layers. Next, we demonstrate the contribution of individual knobs on a full model using the GA-based obfuscator. We let only one obfuscating knob be available at a time during the GA search, and keep a fixed budget of $B= 0.02$. The results are shown in~\cref{fig:SQ_individual}. Among all four sequence obfuscating knobs, layer branching and selective fusion have higher LER for the same latency budget and are clearly better choices. 
The selective combination of 4 knobs by NeurObfuscator achieves stronger obfuscation than any single knob, as expected.

\begin{table}[htbp]
\caption{Effect of Individual Sequence Obfuscating Knobs}
\begin{center}
\centering
\resizebox{0.98\linewidth}{!}{
\begin{tabular}{ ccc } 
 \toprule
 
 \textbf{Knobs} & \textbf{Extra Operator} & \textbf{Latency Overhead}\\ 
 \midrule
 Branching (output-wise by 2) & 1 $\times$ Conv2D, 1 $\times$ Concate &21\% $\sim$ 27\% \\ 
 \midrule
 Branching (output-wise by 4) & 3 $\times$ Conv2D, 1 $\times$ Concate &38\% $\sim$ 49\% \\ 
 \midrule
%   Branching (Input-wise by 2) & 1 $\times$ Conv2D, 1 &27\% $\sim$ 32\% \\ 
%  \midrule
%  Branching (Input-wise by 4) & 3 $\times$ Conv2D, 1 &71\% $\sim$ 76\% \\ 
Selective Fusion (N=0)& 1 $\times$ ReLU, 1 $\times$ BN &14\% $\sim$ 15\% \\ 
 \midrule
 Deepen& 1 $\times$ Conv2D (1x1 kernel) &39\% $\sim$ 89\% \\ 
 \midrule
 Skipping & 1 $\times$ Conv2D &70\% $\sim$ 130\% \\ 
%  Stride & 0.027 & 0.000 & 0.000 & 0.007 & 0.0007 & 0.0004  \\
 \bottomrule
\end{tabular}
}
\label{tab:individual_knobs}
\end{center}
\vspace{-5pt}
\end{table}

% At the end, complex model such as ENAS-net and MobileNet-V2 are dicussed, though the predictor fails to predict their layer sequence correctly even without obfuscation.

% \jtnote{Replace the MobileNet results with a random generated architectures to validate.}
% For the NAS architecture example, we select a CNN model derived using ENAS \cite{pham2018efficient} on CIFAR-10 and a model derived by DenseNAS \cite{fang2020densely} on ImageNet. \jtnote{DenseNAS has not been added yet. Should use DenseNAS -> ResNas version for both ImageNet and CIFAR-10.}

\begin{figure}[htbp]
    \centering
    \includegraphics[width=1.0\columnwidth]{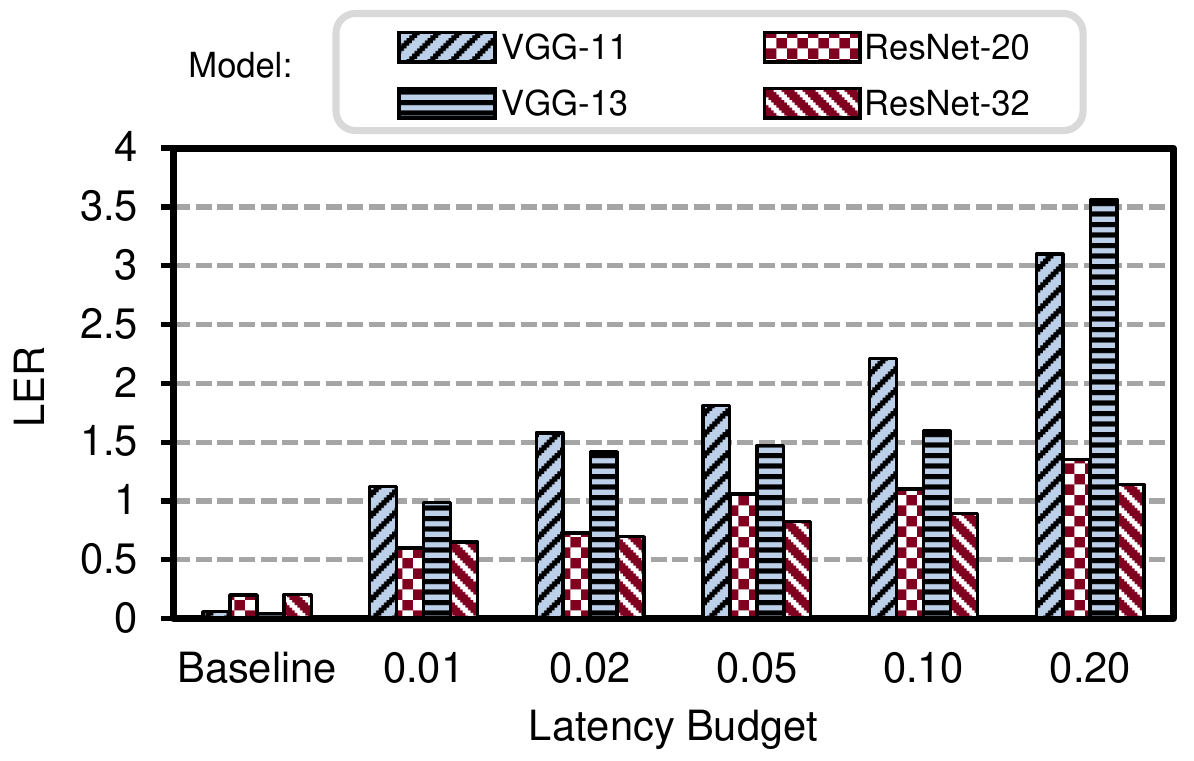}
    \caption{Sequence obfuscation results on typical architectures on CIFAR-10 dataset including VGG-11, VGG-13, ResNet-20 and ResNet-32. }
    \label{fig:obfuscate1}
\end{figure}

\textbf{NeurObfuscator - Sequence Obfuscation}. 
%NeurObfuscator uses a genetic algorithm to search for the best combination of the four sequence obfuscating knobs.
We demonstrate the performance of NeurObfuscator on CIFAR-10 and ImageNet datasets. For VGG-11, VGG-13 and ResNet-32 running on CIFAR-10,
we use bagging of all 15 LSTM predictors. 
The LER results  under different latency budgets are shown in~\cref{fig:obfuscate1}.
We notice that the LER absolute value is high for VGG-11 and VGG-13 while low for ResNet-20 and ResNet-32.
This is because LER is the layer editing distance divided by total number of layers of the vanilla architecture (without obfuscation) and the sequence obfuscation affects the absolute editing distance directly rather than the relative editing distance (i.e., LER). 

\begin{figure}[htbp]
    \centering
    \includegraphics[width=1.0\columnwidth]{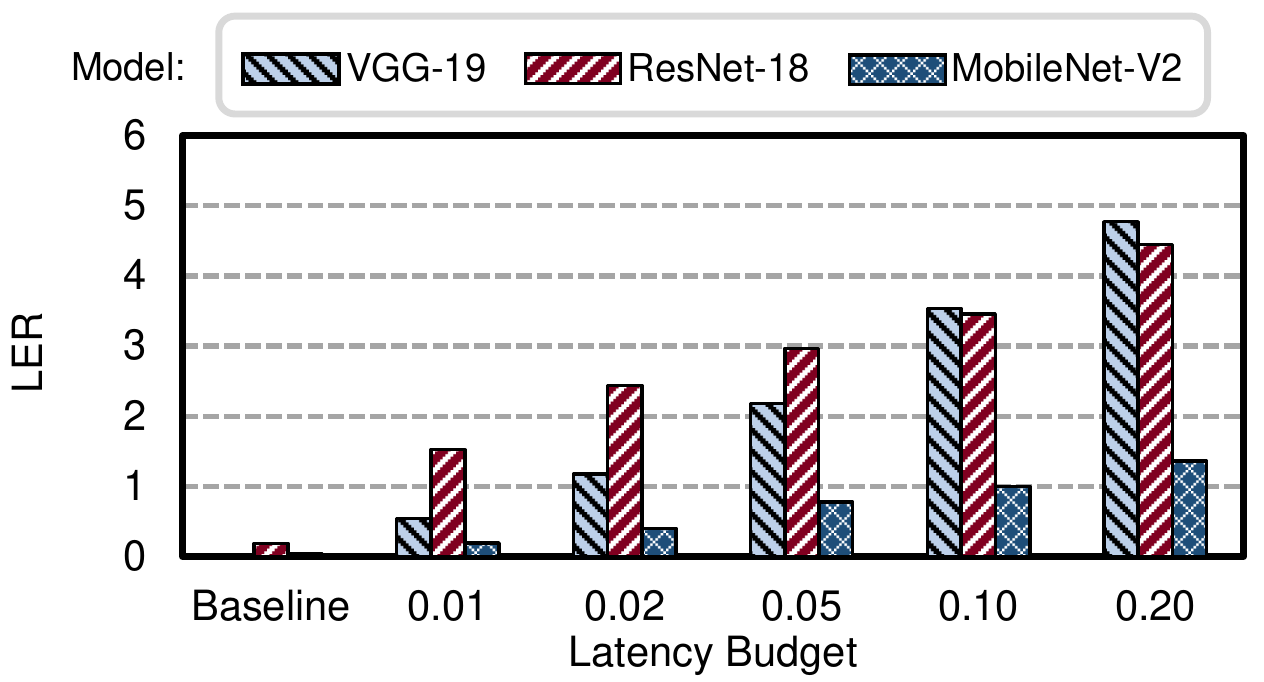}
    \caption{Sequence obfuscation results on typical architectures on ImageNet dataset including VGG-19, ResNet-18 and MobileNet-V2.}
    \label{fig:obfuscate2}
\end{figure}

For VGG-19, ResNet-18 and MobileNet-V2 on ImageNet dataset, case A and case B LSTM predictors struggle to get good extraction performance, i.e., provide low LER for the baseline architecture.  So we use bagging of three ``elite'' LSTM predictors (number of units of 128, 256, 512 LSTM predictors in case C), which have near-zero clean LER. The results are shown in~\cref{fig:obfuscate2}. 
Moreover, we observe that LER increases sub-linearly with increasing latency budget. This is because the search space is kept fixed, and the most effective knobs with low latency overhead are chosen up front and so increasing the budget only allows knobs that are not as effective to get added to the obfuscation set.

\textbf{Summary 1:}
We demonstrate the performance of four knobs, namely, layer deepening, layer skipping, layer branching and selective fusion, for sequence obfuscation. While layer branching and selective fusion have relatively strong performance, combination of all four knobs by GA in NeurObfuscator results in the strongest performance. We evaluated our tool on multiple models taking CIFAR-10 and ImageNet datasets as input data. On a ResNet-18 ImageNet model, we achieved a 2.44 LER (translates to 44 layers' difference) with a mere 2\% inference latency overhead.

\subsection{Dimension Obfuscation Performance}
\begin{figure}[htbp]
    \centering
    \includegraphics[width=\linewidth]{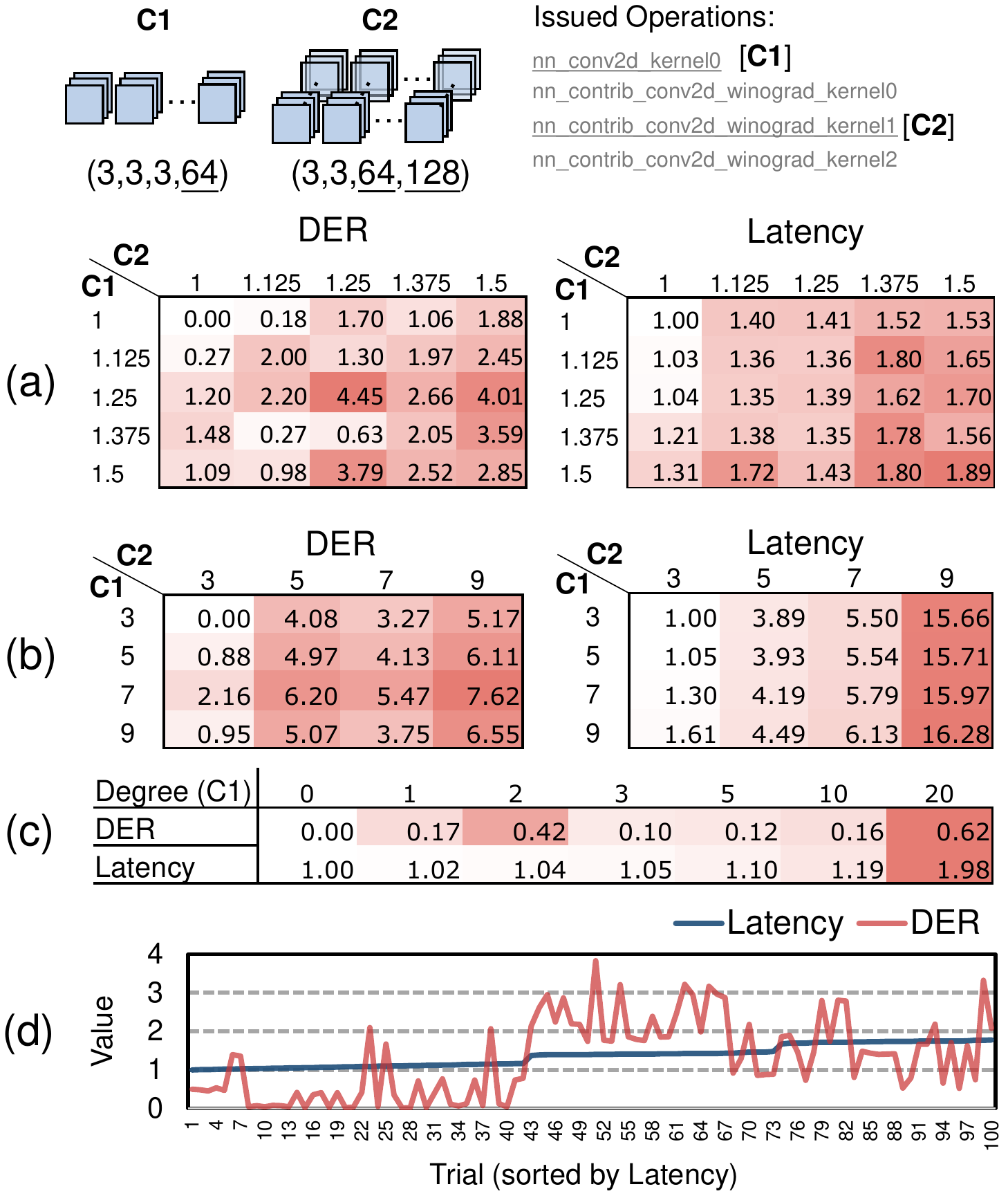}
    \caption{Dimension Obfuscation on a Conv2D layer with 64 input channel and 128 output channel. Application of (a) layer widening to C1/C2, (b) kernel widening to C1/C2, (c) dummy addition to only C2, and (d) random schedule modification that results in highest DER for a given latency overhead.}
    \label{fig:parameter_all}
\end{figure}

For dimension obfuscation, we use the RF regression testbed and DER metric (\cref{eq:DER}) to evaluate the effect of obfuscation.
We pick a Conv2D layer (C2) with 64 input channels and 128 output channels with $3\times3$ kernel from VGG-19 network as an example. The top subfigure in~\cref{fig:parameter_all} shows the layer operators marked by sequence obfuscation. Here  C1 is the input Conv2D layer with 3 input channels and 64 output channels.
In this example, the job of dimension extraction is to correctly predict the number of input channels and output channels of C2. We use the predictor to predict the output channel of  C1 and input/output channel of C2. Here the  ground-truth \underline{64} is predicted twice: once as output channel of C1 and once as input channel of C2. The average is taken if the two predictions do not match.
Next, the effect of individual obfuscating knobs is evaluated. The DER and latency overhead  for each knob are shown in~\cref{fig:parameter_all}.
%where the latency overhead is normalized.

\textbf{Layer Widening}. We use grid-search on applying widening factor of 1 to 1.5 (3/2) for C1 and C2. As shown in~\cref{fig:parameter_all}~(a), generally, applying higher widening factor increases the DER, and increases the latency. We found a sweet point where increasing the C1 output channel size by $1.25\times$ can achieve a 1.20 DER with $1.04\times$ latency.

\textbf{Kernel Widening}. Kernel widening affects both types of Conv2D operator. However, as shown in~\cref{fig:parameter_all}~(b), in most cases the large overhead makes it an expensive option to use in practice. The exception is that increasing kernel size of C1 from $3\times3$ to $5\times5$ results in 0.88 DER with $1.05\times$ latency.

\textbf{Dummy Addition}. Dummy addition does not affect the dimension parameters of C2, because dummy operator are issued after ``winograd\_kernel2'' and will not be fused into kernel1.
However for a standard Conv2D such as C1, dummy addition has a dramatic effect.  As shown in~\cref{fig:parameter_all} (c),  DER increases with increasing dummy addition factor and reaches a sweet point when dummy addition factor is 2; the corresponding DER is 0.42 and latency is $1.04\times$.

\textbf{Schedule Modification}. For the schedule modification knob, we target the schedules of two templates (plain-Conv2D and winograd-Conv2D), with a total of 13 distinct tunable parameters. Since the search space is very large, we perform 100 trials of random choices. 
~\cref{fig:parameter_all} (d) plots DER as a function of increasing latency. We see that there are DER spikes (value larger than 1.0) at trials 7, 23 and 25, even when the  increase in latency is 1\%. Thus, schedule modification is by far the most effective knob in dimension obfuscation.

\textbf{NeurObfuscator - Dimension Obfuscation}. We demonstrate the performance of NeurObfuscator on dimension parameter obfuscation.
%NeurObfuscator uses genetic algorithm to search for the best combination of the four dimension obfuscating knobs. 
Using the same GA setting as in sequence obfuscation, and replacing the LER with DER, we obtain the results shown in~\cref{tab:parameter_obfucation}.
The final results are significantly better than when individual obfuscation knobs are used! A high DER of 2.51 is achieved with only 0.02 latency budget, i.e. a 2\% increase in inference  latency. This corresponds to the case where the original layer with 64 input c  
and 128 output channels is extracted to 207 input and 93 output channels.

% \begin{table}[htbp]
% \caption{GA results for dimension obfuscation \jtnote{This table is for old - 64-256 example}}
% \begin{center}
% \centering
% \resizebox{0.85\linewidth}{!}{
% \begin{tabular}{ ccccccc } 
%  \toprule
 
%  \textbf{Budget} & 0.01 & 0.02 & 0.05 & 0.10 & 0.20  \\ 
%  \midrule
%  \textbf{DER} &1.27 & 1.98 & 2.36 & 2.73 & 3.30  \\ 
%  \midrule
%  \textbf{Prediction} & (145, 256) & (191, 256) & (214, 250) & (215, 350) & (273, 246)  \\ 
% %  \midrule
% %  Stride & 0.027 & 0.000 & 0.000 & 0.007 & 0.0007 & 0.0004  \\
%  \bottomrule
% \end{tabular}
% }
% \label{tab:parameter_obfucation}
% \end{center}
% \end{table}

\begin{table}[htbp]
\caption{GA results for dimension obfuscation}
\begin{center}
\centering
\resizebox{0.98\linewidth}{!}{
\begin{tabular}{ cccccccc } 
 \toprule
 
 \textbf{Budget} & 0.00 & 0.01 & 0.02 & 0.05 & 0.10 & 0.20  \\ 
 \midrule
 \textbf{DER} & 0.00 &2.05 & 2.51 & 2.80 & 3.24 & 3.43  \\ 
 \midrule
 \textbf{Prediction} & (64, 128) & (177, 91) & (207, 93) & (225, 92) & (176, 319) & (141, 413)  \\ 
%  \midrule
%  Stride & 0.027 & 0.000 & 0.000 & 0.007 & 0.0007 & 0.0004  \\
 \bottomrule
\end{tabular}
}
\label{tab:parameter_obfucation}
\end{center}
\vspace{-5pt}
\end{table}

% {\bf DO YOU NEED THIS?}\\
\textbf{Summary 2:}
We demonstrate the performance of  four dimension obfuscating knobs, namely, layer widening, kernel widening, dummy addition and schedule modification. While schedule modification has the strongest performance among all four, NeurObfuscator achieves the best dimension obfuscation, as expected. On an example Conv2D layer with 64 input channels and 128 output channels, RF-based dimension extraction achieves 2.05 DER and 2.51 DER under  1\% and 2\% inference latency overhead, respectively.

\section{Conclusions and Future Work}
To mitigate the neural architecture stealing on GPU devices, we propose NeurObfuscator, a NN obfuscating tool that provides both sequence obfuscation and dimension obfuscation. We propose to use a total of eight obfuscating knobs across scripting, optimization and scheduling phases of a neural network model execution. Application of these knobs affect the number of computations, latency and number of memory accesses, thus altering the execution trace. 
To achieve the best obfuscation performance for a user-defined latency overhead, we leverage the genetic algorithm to identify the best combination of obfuscation knobs. 
% We use a genetic algorithm to choose the best combination of obfuscating
% knobs that maximize the obfuscating effect for a user-defined latency overhead.
%With the testbed simulating the real attacker as evaluator, we use genetic-algorithm to solve an optimization problem with latency constraints to maximize the obfuscating effect. 
%A detailed evaluation indicates that the strongest obfuscation for sequence obfuscation is achieved by layer branching and selective fusion, and for dimension obfuscation by modifying the Xgboost generated schedule.
On a ResNet-18 ImageNet model, sequence obfuscation helps achieve a 2.44 LER (which translates to 44 layers' difference) with merely 2\% latency overhead. 
Similarly, with 2\% latency overhead, dimension obfuscation can achieve 2.51 DER corresponding to the case when a 64 input channel, 128 output channel Conv2D gets extracted as 207 input channel and 93 output channel. 

While the proposed methodology has been designed for GPU devices, we plan to extend this to other hardware substrates, such as FPGAs and ASICs.
%For reconfigurable hardware such as FPGA, we believe the reconfigurability itself can stand for a viable obfuscation. For ASIC, the obfuscation can be extended to design phase and many scheduling obfuscation could be possible.
Furthermore, we plan to evaluate the end-to-end performance of such a system. Examples include accuracy evaluation if the obfuscated architecture is used to train a new model, and the attack success rate of the transfer adversarial attack if the obfuscated architecture is used to built a surrogate model.

\bibliographystyle{IEEEtran}
\bibliography{IEEEtran}

\end{document}